\documentclass[]{pasj02} 
\usepackage[switch,mathlines]{lineno} 

\jyear{2024}
\Received{}
\Accepted{}

\usepackage{newtxtext,newtxmath}

\usepackage{graphicx}	
\usepackage{amsmath}	

\newcommand{\Om}{\Omega}	

\newcommand{\om}{\omega}   
\newcommand{\omN}{\omega_{\rm N}}

\newcommand{\omoL}{\omega_{\rm oL}}  
\newcommand{\omiL}{\omega_{\rm iL}} 
\newcommand{\omoF}{\omega_{\rm oF}}  
\newcommand{\omiF}{\omega_{\rm iF}} 

\newcommand{\OmF}{\Omega_{\rm F}}
\newcommand{\OmFm}{\Omega_{\rm ZF}}  

\newcommand{\epsGTE}{\epsilon_{\rm EX}}		
\newcommand{\vre}{\varrho_{\rm e}} 
\newcommand{\Psic}{\Psi_{\rm c}}   
\newcommand{\Psib}{\bar{\Psi}}
\newcommand{\Psiz}{\Psi_0} 
\newcommand{\Psio}{\Psi_1}  
\newcommand{\Psit}{\Psi_2}
\newcommand{\OmH}{\Omega_{\rm H}}
\newcommand{\OmNS}{\Omega_{\rm NS}}  
 \newcommand{\SNS}{S$_{\rm NS}$}
\newcommand{\OLom}{\overline{\om}}	
\newcommand{\SOL}{S$_{\rm oL}$}   
\newcommand{\SIL}{S$_{\rm iL}$}
\newcommand{\SOF}{S$_{\rm oF}$}   
\newcommand{\SIF}{S$_{\rm iF}$}
\newcommand{\Sns}{S$_{\rm NS}$}
\newcommand{\SH}{S$_{\rm H}$}   
\newcommand{\SSN}{S$_{\rm N}$}	 
\newcommand{\Sinf}{S$_{\infty}$} 
\newcommand{\SZAM}{S$_{\rm ZAMD}$}

\newcommand{\calE}{{\cal E}} 
\newcommand{\calEout}{{\cal E}_{\rm (out)}} 
\newcommand{\calEin}{{\cal E}_{\rm (in)}}  
\newcommand{\calEns}{{\cal E}_{\rm NS}}

\newcommand{\calDout}{{\cal D}_{\rm (out)}} 
\newcommand{\calDin}{{\cal D}_{\rm (in)}} 
\newcommand{\calCout}{{\cal C}_{\rm (out)}} 
 \newcommand{\calCin}{{\cal C}_{\rm (in)}} 

\newcommand{\calP}{{\cal P}}
\newcommand{\calPE}{{\cal P}_{\rm E}}  
\newcommand{\calPJ}{\calP_J} 
\newcommand{\calC}{{\cal C}}  
\newcommand{\calR}{{\cal R}}  
\newcommand{\calI}{{\cal I}}  
\newcommand{\vccalI}{\mbox{\boldmath ${\cal I}$} }  
	 
\newcommand{\calPEout}{\calP_{\rm E, (out)}}  
\newcommand{\calPEinU}{\calPE^{\rm (in)}}  
\newcommand{\calPEin}{\calP_{\rm E,(in)}}  
\newcommand{\calPJout}{\calP_{\rm J, (out)}}  
\newcommand{\calPJinU}{\calPJ^{\rm (in)}}  
\newcommand{\calPJin}{\calP_{\rm J,(in)}}  
\newcommand{\calRinf}{{\cal R}_{\infty}}
\newcommand{\calS}{{\cal S}}   
 
\newcommand{\SffH}{\calS_{\rm ffH}} 
\newcommand{\Th}{T_{\rm H}}			
\newcommand{\Sffinf}{\calS_{{\rm ff}\infty}}  
\newcommand{\SN}{\calS_{\rm N}}	 
\newcommand{\SG}{{\cal G}_{\rm N}}  
\newcommand{\GN}{{\cal G}_{\rm N}}  
  
\newcommand{\epsGTEb}{\bar{\epsilon}_{\rm EX}}		
\newcommand{\epsE}{{\varepsilon}_{\rm E}}
\newcommand{\epsJ}{\varepsilon_{\rm J}}	
\newcommand{\SepsE}{S_{\varepsilon_{\rm E}=0}}
\newcommand{\omepsE}{\om_{\varepsilon_{\rm E}=0}}

\newcommand{\OmFb}{\bar{\Omega}_{\rm F}}    

\newcommand{\rH}{r_{\rm H}}

\newcommand{\Dl}{\Delta}  
\newcommand{\Dlom}{\Delta\om}  
\newcommand{\Dlell}{\Delta\ell}  
\newcommand{\al}{\alpha}  
\newcommand{\vp}{\varpi}  

\newcommand{\lb}[1]{\label{eq:#1}} 	
\newcommand{\rf}[1]{\ref{eq:#1}}

\newcommand{\lbf}[1]{\label{ff:#1}} 
\newcommand{\rff}[1]{\ref{ff:#1}}

\newcommand{\beeq}{\begin{equation}} 
\newcommand{\eneq}{\end{equation}}

\newcommand{\benu}{\begin{enumerate}}   
\newcommand{\enen}{\end{enumerate}} 
\newcommand{\vF}{v_{\rm F}}    
\newcommand{\Iot}{I_{\overline{12}}}  
    
\newcommand{\Iout}{I_{\rm (out)}}  
 \newcommand{\Iin}{I_{\rm (in)}}	 
  
 \newcommand{\IinU}{I^{\rm (in)}}
 
\newcommand{\uvt}{\mbox{\boldmath $t$}}

\newcommand{\vcnb}{\mbox{\boldmath $\nabla$}}

\newcommand{\vcj}{\mbox{\boldmath $j$}}  
\newcommand{\vcjp}{\mbox{\boldmath $j$}_{\rm p}}  
 \newcommand{\jt}{j_{\rm t}}
\newcommand{\jvl}{j_{\perp}} 	
\newcommand{\jpl}{j_{\parallel}}

\newcommand{\dr}[2]{\frac{d#1}{d#2}}

 \newcommand{\LPPlDr}[2]{\left(\frac{\partial#1}{\partial#2}\right)}

\newcommand{\LPfrac}[2]{\left(\frac{#1}{#2}\right)}
\newcommand{\DN}[1]{[#1]_{\rm N}}   
\newcommand{\DG}[1]{[#1]_{\rm G}} 
\newcommand{\PN}[1]{(#1)_{\rm N}}    
\newcommand{\PG}[1]{(#1)_{\rm G}}

\newcommand{\kp}{\kappa}

\newcommand{\vcv}{\mbox{\boldmath $v$}}  
\newcommand{\vcm}{\mbox{\boldmath $m$}}

\newcommand{\vcA}{\mbox{\boldmath $A$}}   
\newcommand{\vcell}{\mbox{\boldmath $\ell$}}
\newcommand{\vcvp}{\mbox{\boldmath $v$}_{\rm p}} 
\newcommand{\vvt}{v_{\rm t}}
\newcommand{\vcB}{\mbox{\boldmath $B$}}  
\newcommand{\vcE}{\mbox{\boldmath $E$}}
\newcommand{\vcBp}{\mbox{\boldmath $B$}_{\rm p}}   
 \newcommand{\vcBt}{\mbox{\boldmath $B$}_{\rm t}}
\newcommand{\Bp}{B_{\rm p}} 
\newcommand{\Bt}{B_{\rm t}}
\newcommand{\vcEp}{\mbox{\boldmath $E$}_{\rm p}}	
\newcommand{\OLEp}{\overline{\vcEp}}
\newcommand{\vcS}{\mbox{\boldmath $S$}}
\newcommand{\vcSJ}{\mbox{\boldmath $S$}_{\rm J}}  
\newcommand{\vcSJin}{\mbox{\boldmath $S$}_{\rm J,(in)}}  
\newcommand{\vcSJout}{\mbox{\boldmath $S$}_{\rm J,(out)}}  
\newcommand{\vcSJinU}{\mbox{\boldmath $S$}_{\rm J}^{\rm (in)}}    

\newcommand{\vcSE}{\mbox{\boldmath $S$}_{\rm E}} 
\newcommand{\vcSEout}{\mbox{\boldmath $S$}_{\rm E,(out)}}	
\newcommand{\vcSEin}{\mbox{\boldmath $S$}_{\rm E,(in)}}  

\newcommand{\vcSsd}{\mbox{\boldmath $S$}_{\rm SD}}  
\newcommand{\vcSEM}{\mbox{\boldmath $S$}_{\rm EM}} 

\newcommand{\vcSEMout}{\mbox{\boldmath $S$}_{\rm EM,(out)}}	
\newcommand{\vcSEMin}{\mbox{\boldmath $S$}_{\rm EM,(in)}}

\newcommand{\OLOmFm}{\overline{\Omega}_{\rm ZF}}   
\newcommand{\OLvF}{\overline{v}_{\rm F}} 
\newcommand{\OLvarepsJ}{\overline{\varepsilon}_{J}}    

\newcommand{\OLvcSsd}{\overline{\vcS}_{\rm SD}}
\newcommand{\OLvcSsdin}{\overline{\vcS}_{\rm SD,(in)}}		
\newcommand{\OLvcSsdout}{\overline{\vcS}_{\rm SD,(out)}}   

\newcommand{\OLvcSEM}{\overline{\vcS}_{\rm EM}}  
\newcommand{\OLvcSEMout}{\overline{\vcS}_{\rm EM,(out)}}
\newcommand{\OLvcSEMinU}{\overline{\vcS}_{\rm EM}^{\rm (in)} }    

\newcommand{\vcSEMinU}{\vcSEM^{\rm (in)}}

\newcommand{\barom}{\bar{\om}}

\newcommand{\zam}{zero-angular-momentum} 		
 
\newcommand{\SgapI}{S$_{\rm G(in)} $}  
\newcommand{\SgapO}{S$_{\rm G(out)}$}

\newcommand{\ggel}{ \mathrel{
\raise1.1ex\hbox{$\scriptstyle >$}
\mkern-10mu\raise.4ex\hbox{$\scriptstyle =$}
\mkern-10mu\lower0.3ex\hbox{$\scriptstyle <$} }}
\newcommand{\lleg}{ \mathrel{
	\raise1.1ex\hbox{$\scriptstyle <$}
	\mkern-10mu\raise.4ex\hbox{$\scriptstyle =$}
	\mkern-10mu\lower0.3ex\hbox{$\scriptstyle >$} }}
\newcommand{\ggo}{ \mathrel{\raise.3ex\hbox{$>$}\mkern-14mu\lower0.6ex\hbox{$\sim$}} }
\newcommand{\lo}{\mathrel{\raise.3ex\hbox{$<$}\mkern-14mu\lower0.6ex\hbox{$\sim$}} }

\newcommand{\SiF}{S$_{\rm iF}$}	
\newcommand{\SoF}{S$_{\rm oF}$}
\newcommand{\TTH}{T_{\rm H}} 
 
\newcommand{\RH}{\calR_{\rm H}}

\newcommand{\xN}{x_{\rm N}}	
\newcommand{\xM}{x_{\rm M}}
\newcommand{\xE}{x_{\rm E}}	
\newcommand{\xoL}{x_{\rm oL}}   
\newcommand{\xiL}{x_{\rm iL}}

\newcommand{\SSM}{{S$_{\rm M}$}} 
\newcommand{\SSMc}{{S$_{\rm Mc}$}}  
\newcommand{\SE}{S$_{\rm E}$} 
\newcommand{\omM}{\om_{\rm M}}
\newcommand {\hc}{h_{\rm c}}

\newcommand{\RTDy}{rotational-tangential discontinuity}

\begin{document} 

\title{Electromagnetic Energy Extraction in Kerr Black Holes through Frame-Dragging Magnetospheres}

\author{Isao \textsc{Okamoto},
\altaffilmark{1}%
\email{iokamoto@jcom.zaq.ne.jp}
Toshio \textsc{Uchida},
\altaffilmark{2}%
\email{uchidat@meiji.ac.jp}
and Yoogeun \textsc{Song},
\altaffilmark{3}
\email{ygsong1004@gmail.com}}


\altaffiltext{1}{National Astronomical Observatory, 2-21-1 Osawa, Mitaka-shi,
Tokyo 181-8588, Japan}
\altaffiltext{2}{School of Political Science and Economics, Meiji University, 1-9-1 Eifuku, Suginami-ku, Tokyo, 168-8555, Japan}
\altaffiltext{3}{Mullard Space Science Laboratory, University College London,  Holmbury St.\ Mary, Dorking, Surrey, RH5 6NT, United Kingdom}

\KeyWords{stars: black holes${}_1$ --- acceleration of particles${}_2$ ---  magnetic fields${}_3$ --- methods: analytical${}_4$}

\maketitle

\begin{abstract}
It is argued that the zero-angular-momentum-observers (ZAMOs) circulating with the frame-dragging-angular-velocity $\om$ plays a leading part in energy extraction.  
When the condition $\OmF<\OmH$ is satisfied, where $\OmH$ and $\OmF$ are the horizon and field-line (FL) angular-velocities (AVs), they will see that the null surface \SSN\ with $\omN=\OmF$ always exists in the force-free magnetosphere. The pivotal ZAMO-measured FLAV $\OmFm\equiv \OmF-\om$ changes sign on this surface \SSN\ where the force-free and freezing-in conditions break down. The force-free magnetosphere is divided on this surface, with particle-current sources on it. The outer domain $\calDout$ outside \SSN\ spins forward ($\OmFm>0$), whereas the inner domain $\calDin$ inside spins backward ($\OmFm<0$). 
Because the electric field $\vcEp$ reverses direction there, the Poynting flux also reverses direction from outward to inward, though the positive angular momentum always flows outwardly.
Electromagnetic extraction of energy will be possible only through the frame-dragged magnetosphere, with the inner domain $\calDin$ nested between the horizon and the surface \SSN, when $\OmF<\OmH$ is ensured by the 1st and 2nd laws of thermodynamics.  
\end{abstract}


\section{\label{sec:level1}Introduction}   

More than four decades have passed since the pioneering paper by \citet{bla77} (referred to as BZ77) was published on the electromagnetic extraction of energy from Kerr black holes (BHs).  This complicated and formidable task, however, still remains a significant challenge even in the latest new frontier in modern classical physics \citet{tho17}.  The purpose of this paper is to challenge this task by taking the ill-understood frame-dragging effect fully into account.  

Fundamental concepts and expressions necessary for elucidating extraction of energy from Kerr holes have fortunately been given, as well as the basic formulation of general relativity, thermodynamics and electrodynamics, in the almost complete form already four decades ago (BZ77; \cite{mac82} (MT82); \cite{zna77,zna78,bla79,tho86} (TPM86)). \cite{phi83a,phi83b} developed a comprehensive model for BH-driven hydromagnetic flows or jets for active galactic nuclei (AGNs).   These references presume, however, that a battery exists on the event horizon, with magnetic field lines threading the horizon.  That is, a magnetized Kerr hole would possess not only a battery but also an internal resistance on the horizon. We refer to this model as the single-pulsar model with a single electric circuit. This is because the magnetosphere consists of double wind structures with a negligible violation of the force-free condition in-between for particle production, with a battery on the horizon and two resistances on the horizon and infinity surface in a single series circuit. 

\citet{uch97a,uch97b} constructed a comprehensive theory of force-free electromagnetic fields, where it was pointed out that the breakdown of force-freeness is not a defect proper to the force-free approximation.  \citet{gra14} made extensive use of his theory as a `spacetime approach' to force-free magnetospheres of various objects. \citet{bar21} stated that many references on astrophysical force-free plasma leave open the issue of identifying the actual sources of the force-free electromagnetic field.  We show here that the actual breakdown of the force-free condition we meet with provides rather firm foundations necessary to rebuild an active force-free magnetosphere, including the built-in particle-current sources.

We remake the pulsar force-free magnetosphere, with the frame-dragging effect fully taken into account. The central premise is that the large-scale poloidal magnetic field $\vcBp$ will be trapped (and frame-dragged) by the hole into circulation around the hole with the frame-dragging-angular velocity (FDAV) $\om$.  When each magnetic FL extends from near the horizon \SH\ to the infinity surface \Sinf, with the field-line-angular-velocity (FLAV) $\OmF$ kept constant (Ferraro's law of iso-rotation),  the zero-angular-momentum-observers (ZAMOs) circulating with the FDAV $\om$ will then see that Ferraro's law no longer holds for the ZAMO-measured FLAV $\OmFm=\OmF-\om$ (see equation\ (\rf{twoOmFa})). The force-free magnetosphere consisting of charge-separated plasma is divided by the null surface \SSN\ where $\OmFm=0$, i.e., $\om=\OmF\equiv\omN$, into the two domains, outer semi-classical (SC) and inner general-relativistic (GR), $\calDout$ and $\calDin$. The ZAMOs will soon realize that the breakdown of the force-free and freezing-in conditions is inevitable on the null surface \SSN\ with $\OmFm\ggel 0$ for $\om\lleg\OmF$. 

Making reference to the Membrane Paradigm (TPM86), 
we think of \emph{three} membranes in our twin-pulsar model. The first \emph{two} resistive ones on the infinity and horizon surfaces, $\Sffinf$ and $\SffH$ terminate the two force-free domains $\calDout$ and $\calDin$ by particle acceleration on $\Sffinf$ and entropy production on $\SffH$, respectively. The third \emph{one} is the inductive membrane $\SN$ on the null surface \SSN, which hides the particle-current sources due to the breakdown of force-freeness in the force-free limit. 

 \citet{bla22} recently constructed the model for the ergo-magnetosphere, ejection disc, and magnetopause in M87. They argued that the force-free approximation is justifiable in the vicinity of the black hole.  There is actually good chemistry between thermodynamics for the Kerr hole with \emph{two} hairs (e.g.,\ entropy $S$ and angular momentum $J$) and electrodynamics for the pulsar magnetosphere with \emph{two} conserved quantities along each FL, $\OmF$ and $I$.  This is thanks to the frame-dragging effect, which plays a key role in coupling force-free electrodynamics with BH thermodynamics. We show that the force-free approximation is flexible and robust enough to accept the `breakdown' of the force-free condition (as well as the freezing-in one) on the null surface \SSN, which always exists between the two light surfaces, \SOL\ and \SIL when the hole loses energy (BZ77). 

A brief outline of each section\ is as follows.    
Section\ \ref{ther-rot} points out that the Kerr holes are a rotating mass of `entropy matter', and hence, without invoking the influx of \emph{negative} angular momentum, it will be implausible to produce the outgoing flux of \emph{positive} angular momentum, thereby extracting energy from the Kerr BH. 
In section\ \ref{FFA}, 
from a thermodynamic point of view, we discuss the similarity and dissimilarity of the force-free magnetospheres between Kerr holes and pulsars in sections\ \ref{KBH-M} and \ref{P-M}.  

In section\ \ref{BH-GE},  
in order to elucidate the Kerr hole's elaborate mechanism of energy extraction, we use the wind, flux and circuit analyses expediently. 
In particular, section\ \ref{cir-appror} argues that a pair of unipolar induction batteries with electromotive forces (EMFs), $\calEout$ and $\calEin$  will drive currents to flow through the circuits $\calCout$ and $\calCin$ in the outer and inner domains, $\calDout$ and $\calDin$. 
There will be a huge voltage drop $\Dl V$ across \SSN\ between the two EMFs for particle production. 

Section\ \ref{EnegR}    
shows that the angular-momentum density of the electromagnetic field is positive ($\epsJ>0$) in $\calDout$ and negative ($\epsJ<0$) in $\calDin$. That is, the null surface \SSN\ coincides with the zero-angular-momentum-density surface \SZAM.   The energy density $\epsE$ changes from positive to negative somewhere inward beyond the null surface \SSN. 

In section\ \ref{NullS}, 
the ZAMOs will see that reversal of the electric field $\vcEp$ and the Poynting flux $\vcSEM$ on \SSN\ leads inevitably to the breakdown of the force-free and freezing-in conditions. We show how critical the resulting severance of the current- and stream-lines on \SSN. For example, this surface \SSN\ will be widened to such a gap $\GN$ as filled with zero-angular-momentum-particles (ZAM-particles) pair-produced due to the voltage drop $\Dl V$ (see section \ref{cir-appror}).

In section\ \ref{m-mdGap},  
the structure of a gap $\GN$ under the inductive membrane $\SN$ is discussed with respect to the pinning-down of threading FLs on the pair-created plasma, resulting in magnetization of it and a magneto-centrifugal plasma-shed on \SZAM, etc. (see also figure \rff{GapI} for a simple model for the current function $I$ for a typical FL). 

In section\ \ref{BC-SN},  
we argue the boundary condition for determining the eigenvalue of $\OmF=\omN$ on \SSN\ in the steady axisymmetric state. Some new pieces of evidence shown will be helpful in understanding an enigmatic flow of \emph{positive} angular momentum from the horizon membrane $\SffH$, beyond the inductive membrane $\SN$ covering the Gap $\GN$, to the infinity membrane $\Sffinf$ (see section\ \ref{BCagain}). Thus, the ZAM-Gap $\SG$ will allow us to impose the conservation law of angular momentum as the boundary condition determining the eigenfunction $\OmF(\Psi)=\omN$; this means that the eigen-magnetosphere with $\omN=\OmF$ is `frame-dragged' by the hole's rotation (see section\ \ref{Feigenv}). 

Section\ \ref{TW-P-M}  
attempts to explain the null surface \SSN\ in terms of a new kind of \RTDy\ (RTD) in the GR setting \citep{lan84,oka15a}. We conjecture that this RTD involving a voltage drop $\Dl V$ between the two EMFs will bring up a new mechanism of pair-particle creation at work on \SSN\ toward widening to a GR gap $\GN$. As opposed to the single-pulsar model based on BH electrodynamics with a negligible violation of the force-free condition, we propose the twin-pulsar model based on GTED because there will be `two pulsar-type magnetospheres' coexisting, outer prograde- and inner retrograde-rotating, respectively, with the RTD spark-Gap in-between for the supply of current-particles.   

Section\ \ref{FluxC} 
discusses the energetics and structure of the twin-pulsar model as opposed to that of the single-pulsar model (cf.\, e.g.\ IV D in TPM86.)     
The last section\ \ref{Dis-Con} 
is devoted to discussions and conclusions, with some remaining issues listed. 

Appendix A discusses the approximate position-shape of the null surface \SSN\ in the force-free magnetosphere for the spin parameter $h=a/\rH$ (see equation\ (\rf{Khole-h})).

\section{The Kerr black hole as a thermodynamic object}  \label{ther-rot}  
\setcounter{equation}{0}
\numberwithin{equation}{section}

The no-hair theorem indicates that Kerr BHs possess only two hairs.  When one chooses $S$ and $J$ as two extensive variables, then all other thermodynamic quantities are expressed as functions of these two. For example, the BH's mass-energy $c^2 M$ is expressed in terms of $S$ and $J$, as follows;   
\beeq 
M=\sqrt{(\hbar cS/4\pi kG) + (\pi kcJ^2/\hbar GS)}.  
\lb{massF} \eneq  
As one can, in principle, utilize a Kerr BH as a Carnot engine \citep{KO91}, it may be regarded as a thermodynamic object but not as an electrodynamic one because the Kerr BH will by itself store no extractable electromagnetic energy. Therefore, the Kerr BH should be regarded rather as a huge rotating `mass of entropy matter' (see, e.g., \cite{bas90}), fundamentally different from the magnetized rotating NS, which consists of `normal matter' with magnetic FLs emanating outside. Thus, its evolutionary behaviours, such as due to the extraction of angular momentum, are strictly governed by the four laws of thermodynamics (see, e.g., \cite{tho86} for a succinct summary). 

The mass $M$ of the hole is divided into the irreducible and rotational masses, i.e.,
\begin{subequations}
\begin{eqnarray}
M=M_{\rm irr}+M_{\rm rot},  \hspace{0.3cm}~~~~~ \lb{massa} \\  
M_{\rm irr}=\frac{M}{\sqrt{1+h^2}} =\sqrt{c^4 A_{\rm H}/16\pi G^2} =\sqrt{\hbar cS/4\pi kG}, \hspace{0.3cm} ~~~~~ \\ 
M_{\rm rot}= M[1- 1/\sqrt{1+h^2}], \hspace{0.3cm}  ~~~~~ 	
\end{eqnarray}  \lb{mass/irr/red}   \end{subequations}  
where $A_{\rm H}$ is the horizon surface area, and $h$ is defined as the ratio of $a\equiv J/Mc$ to the horizon radius $\rH$, i.e., 
\beeq 
h=\frac{a}{\rH} =\frac{2\pi kJ}{\hbar S}=\frac{2GM\OmH}{c^3}.  
\lb{Khole-h} \eneq  
The Kerr BH's thermo-rotational state is uniquely specified by $S$ and $J$, or its $M$ and $h$.  We see $h=0$ for a Schwarzschild BH and $h=1$ for an extreme-Kerr BH \citep{OK90,OK91,oka92}. 
The evolutional state of the BH losing energy is then specified as the timeline of the function $h(t)$ for the `outer horizon' in $0\lo h\lo 1$. 

The BH's $M_{\rm irr}$ and $A_{\rm H}$ are functions of $S$ only, but $M_{\rm rot}$ may be a function of $S$ as well as $J$. Therefore, only when the hole loses angular momentum (i.e., $dJ<0$), the hole's total mass and rotational mass will decrease, i.e., $dM<0$ and $dM_{\rm rot}<0$, while $dM_{\rm irr}>0$ and $\Th dS>0$ must always hold by the 2nd law. 

Different from a magnetized NS consisting of `normal matter', a Kerr hole with $M=M(S,J)$ in equation\ (\rf{massF}) will be the biggest rotating mass of `entropy matter'. Then, a naive question comes to mind: ``How do magnetic field lines manage to thread and survive in `entropy matter' under the horizon?''   If a battery really existed on the horizon, this might indeed necessitate the threading of FLs into the matter under \SH\ such as, e.g., the `imperfect conductor' (BZ77; \cite{zna78}).  

The zeroth law of thermodynamics indicates that two `intensive' variables, $\Th$ (the surface temperature) and $\OmH$, conjugate to $S$ and $J$, respectively, are constant on \SH, e.g., $\om\to\OmH$ for $\al\to 0$.  In passing, the third law indicates that ``by a finite number of operations, one cannot reduce the surface temperature to the absolute zero with $h=1$.''  In turn, ``the finite processes of mass accretion with angular momentum cannot accomplish the extreme Kerr state with $h=1$, $\Th=0$ and $\OmH=c^3/2GM$'' \citep{OK91}.
Incidentally, the `inner-horizon' thermodynamics can formally be constructed analogously to the `outer-horizon' thermodynamics \citep{OK92, cve18}. 

It is the first and second laws that govern the extraction process of energy, i.e., 
\begin{subequations}  \begin{eqnarray}	  \hspace{2cm}
c^2 dM=\Th dS + \OmH dJ ,  \lb{1st-l} \\  
\Th dS \geq 0 ,  
\end{eqnarray} \lb{1-2laws}   \end{subequations}  
where $\Th$ and $\OmH$ are uniquely expressed in terms of $J$ and $S$ from equation\ (\rf{massF}) or $M$ and $h$ \citep{OK90}; 
\beeq 
\Th=c^2 (\partial M/\partial S)_J, 
\quad     
\OmH=c^2 (\partial M/\partial J)_S.  
\lb{2nd-l}   \eneq  

When Kerr BHs are regarded as a substantial rotating mass of entropy matter confined by its self-gravity within the event horizon, we conjecture that they will not allow the presence of magnetospheric field lines anchored in the matter inside the horizon, and hence the Kerr BH itself will be unable  
to behave like a battery (see \cite{pun89}). 

It will nevertheless be argued here that Kerr BHs can acquire and keep a force-free magnetosphere by making full use of frame-dragging   
(see \cite{oka92,oka15a}; references therein). 



\section{The force-free approximation and BH thermodynamics}   \label{FFA} 
\setcounter{equation}{0}
\numberwithin{equation}{section}  

\subsection{Kerr black hole force-free magnetospheres} \label{KBH-M} 
 
The magnetospheres filled with perfectly conductive plasma around NSs and Kerr BHs are considered under the two basic presumptions of the force-free and freezing-in conditions in the stationary axisymmetric state with $\vcE_{\rm t} =0$. These two conditions are given by  
	\begin{subequations}	\begin{eqnarray}  
	\vre\vcE+\vcj /c \times \vcB=0, \lb{ff-fz}  \\  
 \vcE+\vcv/c\times\vcB=0,  \lb{fi-c}  
	\end{eqnarray} \lb{ff-a}  \end{subequations}  %
where $\vcB$ and $\vcE$ are the electromagnetic fields, $\vcj$ is the electric current and $\vcv$ is the velocity of force-free charged particles. All electromagnetic quantities are measured by the ZAMOs circulating with the FDAV $\om$. The magnetosphere is then characterized by the two quantities, $\OmF(\Psi)$ and $I(\Psi)$. The latter is the angular-momentum flux/current function. Both are conserved along each FLs, 
where $\Psi$ is the stream function, and $\OmF$ and $\om$ are the AVs relative to absolute space (MT82). Then, we define the FLAV referred to as the ZAMOs\footnote{denoted by $\Om_{{\rm F}\om}$ when defined for the first time by equation (5) in \citet{oka15a}.}; 
    \begin{subequations}  \begin{eqnarray} 	
	\OmFm=\OmF - \om.		\lb{twoOmFa} 
    \end{eqnarray}  
Then, we can express the conserved FLAV $\OmF$ in terms of the two non-conserved AVs; 
 \begin{eqnarray} 	
	\OmF= \OmFm + \om.		\lb{twoOmFb}  
	\end{eqnarray} \lb{twoOmF}   \end{subequations}     
We emphasize the need to define $\OmFm$ ``additionally". This is related to the statement \citep{pun90} that ``... the present solutions for the BZ process are incomplete and that additional physics is needed in order to determine the efficiency of energy extraction." We argue that ``additional and essential" physics needed must be thermodynamics to be unified with electrodynamics with the help of the frame-dragging effect, that is, relations (\rf{twoOmFa},b). 
The $3+1$-formulation of BH electrodynamics by MT82 
was perfectly accomplished analytically, but somehow lacks {\em the finishing touches}, that is, the ZAMO-measured FLAV $\OmFm$ was missing in BZ77, MT82 and TPM86, although the null surface \SSN\ was already defined by $\omN=\OmF$ (i.e., $\OmFm=0$) in BZ77 (see section \ref{2Flux}).

The null surface \SSN, where $\OmFm\ggel 0$ for $\om\lleg\OmF$, divides the magnetosphere unequivocally into the two domains, i.e., the outer SC and inner GR domains. The ZAMO-FLAV 
$\OmFm$ is also related to the Poynting flux (see equations (\rf{Sem}) and (\rf{dS},b)), and hence plays an indispensable role in defining the efficiency of energy extraction by the 2nd law of BH thermodynamics, as already shown in BZ77. 
Equations (\rf{twoOmFb},a) will lead to the 1st and the 2nd laws, respectively, on \SH\ where $\al\to 0$ and $\om\to \OmH$ (see equations (\rf{Laws-(ii)}a,b)). 

The poloidal and toroidal components of $\vcB$, the electric field $\vcEp$ and the ZAMO-FLRV (field-line-rotational-velocity) $\vF$ are defined by 
\begin{subequations}		\begin{eqnarray}  	
\vcBp=-(\uvt\times\vcnb\Psi/2\pi \vp), \ \ \vcBt=-(2I(\Psi)/\vp c\al)\uvt , \hspace{3mm} \lb{vcBp/Bt}	\\[1mm]  
\vcEp=-(\OmFm/2\pi\al c)\vcnb\Psi , \hspace{3mm}	\lb{ZAMO-Ep}  	 \\[1mm]  	
\vF=\OmFm\vp/ \al,\hspace{3mm}	\lb{ZAM0-vF}  
\end{eqnarray} \lb{e-m-fZAMO}  \end{subequations} 
where $\uvt$ is the unit toroidal vector and $\vcBt$ is regarded as the swept-back component of $\vcBp$ by inertial loading (i.e., particle acceleration and entropy production; see section  \ref{Iout/in}).  Because $\vF$ stands for the physical velocity of FLs relative to the ZAMOs, $\vcEp$ is entirely induced by the motion of the magnetic field lines, i.e., $\vcEp =-(\vF\uvt/c)\times \vcBp$. 

The role of the FDAV $\om$ is thus to make the ZAMO-FLAV $\OmFm$ `violate' Ferraro's law along each FL and change sign on \SSN, and hence $\vcEp$ in equation\ (\rf{ZAMO-Ep}) changes direction as well.  This leads to the elucidation of `how and where' the basic conditions given in equations\ (\rf{ff-a}a,b) should be broken down, thereby creating the current-particle sources and opening the unique and unequivocal path toward GTED (see section\ \ref{BH-GE}). 

The AV $\OmF$ possesses the two sides, i.e., the FLAV and the electric potential gradient. With respect to $I$, `current/angular-momentum duality' also holds in the degenerate state.  From equations (5.6a,b,c) in MT82 
for the current and charge densities, we have the poloidal and toroidal components of $\vcj$ and the charge density $\vre$; 
	\begin{subequations}  \begin{eqnarray} 	
	\vcjp=-\frac{1}{\al}\dr{I}{\Psi}\vcBp,		\lb{vcj-p}  \\[1mm]  
	\jt=\vre\vF + \frac{2 I}{\vp\al^2 c}\dr{I}{\Psi},   \lb{vcj-t} \\[1mm]  
	\vre= -\frac{1}{8\pi^2 c} \nabla\cdot \left( \frac{\OmFm}{\al}\vcnb\Psi  \right). \lb{bh/rho-e}   
	\end{eqnarray} \lb{vcj-rhoe}   \end{subequations}
Because the hole's gravity produces a gravitational redshift of ZAMO clocks, their lapse of proper time $d\tau$ is related to the lapse of global time $dt$ by the lapse function $\al$, i.e., $d\tau/dt=\al$ (see MT82). 

The angular momentum and energy fluxes are given in terms of two conserved quantities $\OmF(\Psi)$ and $I(\Psi)$; 
	\beeq
	\vcSE =\OmF(\Psi)\vcSJ, \quad \vcSJ=( I(\Psi)/2\pi\al c)\vcBp,  
	\lb{E-AmFlux} \eneq		
where the toroidal component of $\vcSE$ (and other fluxes) is and will be omitted throughout the paper. 
These are apparently the same as the equations for the pulsar magnetosphere except for the redshift factor $\al$. The output power $\calPE$ and the loss rate of angular momentum $\calPJ$ observed by distant observers are given by
\begin{subequations}  \begin{eqnarray}  
	\calPE= -c^2 \dr{M}{t} = \oint \al\vcSE\cdot d\vcA  
= \frac{1}{c}\int^{\bar{\Psi}}_{\Psi_0}\OmF(\Psi) I(\Psi) d\Psi,  	\lb{TotFluxE}  \\	
	\calPJ =-\dr{J}{t} = \oint \al\vcSJ \cdot d\vcA  = \frac{1}{c}\int^{\bar{\Psi}}_{\Psi_0} I(\Psi) d\Psi  
	\lb{TotFluxJ} 	
	\end{eqnarray} \lb{TotalFa,b}  \end{subequations}
(see equations (3.89) and (3.90) in TPM86), where $\vcBp\cdot d\vcA=2\pi d\Psi$ and the integration is done over all open field lines in $\Psi_0\leq\Psi\leq \bar{\Psi}$.  We define the `overall' potential gradient, calculated from $\OmF(\Psi)$ weighted by $I(\Psi)$, i.e.,  
\beeq \left.
\OmFb = \int^{\bar{\Psi}}_{\Psi_0} \OmF (\Psi) I(\Psi) d\Psi \right/ \int^{\bar{\Psi}}_{\Psi_0} I(\Psi) d\Psi=\calPE/\calPJ,    
\lb{GTEepsb}   \eneq  
and then from equations\ (\rf{TotalFa,b}a,b) and (\rf{1st-l}) we have                        
	\begin{subequations} 	\begin{eqnarray}    \hspace{1cm}
		c^2 dM=  \Th dS + \OmH dJ= \OmFb dJ,   \lb{1st-(ii)} \\[1mm]   \hspace{1cm}   
		\Th dS= -(\OmH-\OmFb) dJ ,  \lb{2nd-(ii)}		 
	\end{eqnarray} \lb{Laws-(ii)}    \end{subequations} 
which corresponds to equations (\rf{twoOmF}b,a). 
Then, the 2nd law $\Th dS>0$ requires inequality $\OmH>\OmFb$ to be fulfilled always for the hole to lose energy, i.e., $c^2 dM=\OmFb dJ<0$. 
This situation indicates that the null surface \SSN\ with $\omN=\OmF\approx \OmFb$ always exists (BZ77). This means that the outer SC domain  $\calDout$ ($\OmFm>0$) rotates forward, whereas the inner GR domain $\calDin$ ($\OmFm<0$) rotates backward. 
Thus, the ZAMOs will see that each FL with the same $\OmF(\Psi)$ counter-rotates between the two domains, $\calDout$ and $\calDin$. 

\subsection{Thermodynamics of pulsar force-free magnetospheres} \label{P-M} 
It will be instructive to refer to the \emph{adiabatic-dragging} of the pulsar force-free magnetosphere. The power and the rate of angular momentum loss in equations\  (\rf{TotalFa,b}a,b) are surely applicable to an NS with the surface AV $\OmNS$. The Poynting and angular-momentum fluxes are always outwards, and there is no necessity for reversal of the direction. The FLAV $\OmF$ is uniquely given by the boundary condition for FLs emanating from the surface \Sns\ of a magnetized NS, i.e., 
\beeq
 \OmF=\OmNS. 
\lb{b-cSNS}  \eneq  
The current function $I(\Psi)$ is specified by the criticality condition on the fast magneto-sonic surface S$_{\rm F}$ near infinity, i.e., 
	\beeq
	I_{\rm NS}(\Psi)=\frac{1}{2}\OmF(\Bp\vp^2)_{{\rm ff}\infty}	
		\lb{I/Pul}  \eneq  
(see equation\ (\rf{Iout/in-a})), which is equivalent to Ohm's law for the surface current on the force-free infinity surface $\Sffinf$(\,$\lo$\,\Sinf). The total power is given by simply putting $\OmF=\OmNS$ and $I=I_{\rm NS}$ in equation\ (\rf{TotFluxE}) 
\beeq
\calP_{\rm E: NS} =  \frac{1}{c}\int^{\bar{\Psi}}_{\Psi_0}\OmNS I_{\rm NS} d\Psi 		
=\frac{1}{2c} \int_{ {\cal S}_{{\rm ff}\infty } } \OmF^2(\Bp\vp^2)_{{\rm ff}\infty} d\Psi . 
\lb{Sffinf-NS}  \eneq 		
The related EMF of the NS's surface battery is expressed in terms of potential gradient $\OmF$ by 
\beeq
\calEns=-\frac{1}{2\pi c}\int^{\Psi_2}_{\Psi_1}\OmF(\Psi)d\Psi 
\lb{nsEMF} \eneq	 
\citep{lan84,oka12b},  
which drives currents along FL $\Psit$ with $\vcjp>0$ and return currents along $\Psio$ with $\vcjp<0$, where $\Psiz<\Psio<\Psic<\Psit<\bar{\Psi}$ and $\vcjp\lleg 0$ for $\Psi\lleg\Psic$ (see the outer half of figure \rff{DC-C}; equations (\rf{c-c-c}) and (\rf{vcjpPM})),  where $\bar{\Psi}$ is the last limiting field line satisfying $I(\Psiz)=I(\bar{\Psi})=0$ (see  figure 2 in \cite{oka06} for one plausible example of $I(\Psi)$). The surface return currents flow from $I(\Psit)$ to $I(\Psio)$, crossing FLs between $\Psio$ and $\Psit$ on the resistive membrane $\Sffinf$, and the ohmic dissipation of surface current there formally represents particle acceleration taking place in the resistive membrane $ \Sffinf$ with resistivity $\calRinf$ (see equation\ (\rf{Sffinf-M}) later). That is, the conversion of field energy to kinetic energy takes place on $\Sffinf$ in the form of the MHD particle acceleration \citep{oka74,oka78}. 

Now, we regard the toroidal field $\Bt=-(2I/\vp c)$ as the swept-back component of $\vcBp$ due to inertial loadings on the terminating surface $\Sffinf$ of the force-free domain. Thus, the behaviour of $I=I(\ell, \Psi)$ from the stellar surface to infinity will be described in the pulsar force-free magnetosphere as follows; 
\begin{eqnarray} 
		I(\ell,\Psi) =  \left\{
	\begin{array} {ll} 
		0 & ;\ell \lo \ell_{\rm NS},\ {\rm (no\ resistance)},  \\[1mm]
		I_{\rm NS}(\Psi) &;\ \ell_{\rm NS} \lo \ell \lo \ell_{\rm F},  \ {\rm (FF\ region)}, \\[1mm]
		\to 0 & ; \ell_{\rm F} \lo \ell \lo \ell_{\infty} \ {\rm (particle\ acceleration)}
	\end{array}  \right.    \lb{I/Pul-M}  \end{eqnarray}    
(see equation\ (\rf{OL-I}) for a Kerr hole's force-free magnetosphere). 
We assume simply here that $I(\ell,\Psi)$ approaches zero for $\ell\to\infty$ or $\vp\to\infty$ along each FL. This means that all the Poynting energy is transferred eventually to the particle kinetic energy (see equation\ (\rf{DivvcSE}) later).  
 
NSs of `normal matter' are regarded as innately magnetized, and hence, their magnetospheric FLs will be anchored in the surface layer or crust.  When we think of the 1st law in pulsar-thermo-electrodynamics analogously to the BH case, i.e., 
	\beeq
	c^2 dM_{\rm NS}=T_{\rm NS}dS+\Om_{\rm NS}dJ, 
		\lb{1stNS}  \eneq  
we have $c^2 dM_{\rm NS}=\OmFb dJ$  by equation\ (\rf{GTEepsb}), for energy loss through the force-free magnetosphere. Then, for any entropy generation in the boundary layer, we have 
   \[ T_{\rm NS}dS=-(\Om_{\rm NS}- \OmFb)dJ. \] 
When $\OmF\approx \OmFb=\Om_{\rm NS}$, the boundary condition (\rf{b-cSNS}) ensures $T_{\rm NS}dS=0$, that is, no entropy production near the boundary layer on \Sns. This will surely ensure {\em adiabatic} extraction of energy from the NS through its force-free magnetosphere. There is naturally no necessity for the breakdown of the force-free condition anywhere (except but a negligible violation for a shortage of particles). In other words, the NS can `adiabatically' be dragging the force-free magnetosphere, through which the NS blows the magneto-centrifugal wind outward in the form of a Poynting flux. 

On the other hand, Kerr BHs will consist of a huge mass of `entropy matter' confined within the event horizon, and there seems to be no theoretical nor observational evidence so far indicating that they are capable of being magnetized nor anchoring the magnetospheric FLs. When the adiabatic extraction of energy is excluded and  $\OmH>\OmF$ is the case, the ZAMOs will notice that the null surface \SSN\ where $\OmFm=0$ always exists between the inner and outer domains, $\calDin$ and $\calDout$, counter-rotating each other, and the force-free condition must break down there to establish a pair of batteries with a voltage drop in-between to construct the particle-current sources (see section\ \ref{NullS}).

\section{Toward BH gravito-thermo-electrodynamics}  \label{BH-GE}   
\setcounter{equation}{0}
\numberwithin{equation}{section}  

 In order to clarify the fundamental properties of the force-free magnetosphere of Kerr BHs, we expediently use wind, flux, and circuit analyses, which must be complementary to each other. The key point is to keep the ZAMOs', i.e., the physical observer's viewpoint, whose index is given by the ZAMO-FLAV $\OmFm$ in equations\ (\rf{twoOmFa},b). 
 
\subsection{The wind flow analysis}   \label{Wind-A}  
\subsubsection{The velocity $\vcv$ of `force-free' particles } \label{velocityP}  
  Combining the two conditions (\rf{ff-a}a,b), with use of equations\ (\rf{vcj-rhoe}a,b,c) for $\vcj$ and $\vre$, we have $\vcv=\vcj/\vre$ for the velocity of the force-free particles
  	\beeq  
	\vcv=\frac{\vcj}{\vre}= -\frac{1}{\vre\al}\dr{I}{\Psi}\vcB+\vF \uvt , 
	\lb{vc-vjA} \eneq 
which indicates that FLs, current- and stream-lines (FCSLs) are thus parallel to each other and must be equipotential in the force-free domains. And yet, the force-free plasma must be charge-separated (BZ77), and the role of force-free particles is just to carry charges, exerting no dynamical effect.   

On the other hand, the axial symmetry $\vcE_{\rm t}=0$ imposes $\vcjp\times\vcBp=0$ and $\vcvp\times\vcBp=0$, respectively, in equations (\rf{ff-a}a,b), and hence $\vcjp=\eta\vcBp$ and $\vcvp=\kappa\vcBp$, i.e., $\vcjp=\vre\vcvp=\vre\kappa\vcBp\equiv \eta\vcBp$, 
where $\eta$ and $\kappa$ are a scalar function of the position along each FL. 

The ZAMO-measured particle-velocity $\vcv$ is summarized as follows;  
	\begin{subequations} \begin{eqnarray}
		\vcv=\kappa\vcB+\vF \uvt, \lb{vcv} \\  
		\vcvp=\kappa \Bp, \quad \vvt=\kappa\Bt +\vF, \lb{vcvp/t} \\ 
 	\kp=-(1/\vre\al) (dI/d\Psi),  \lb{kappa}  
\end{eqnarray}   \end{subequations}   
where $\vF$ is given by equation\ (\rf{ZAM0-vF}).

\subsubsection{The null surface \SSN\ between two light surfaces, \SOL\ and \SIL }  \label{SoL/SiL} 
When the ZAMOs see along each FL in case of $\OmH>\OmF>0$ from the horizon to infinity, they will find that $\OmFm$ increases linearly with $\om$ from $-(\OmH-\OmF)$ on \SH, beyond null on \SSN\ where $\om=\omN=\OmF$, to $\OmF$ on \Sinf\ on the \emph{same} FL, because $\om$ decreases from $\OmH$ on \SH\ to zero on \Sinf.  Thus, by putting $\vF=\mp c$ in equation\ (\rf{ZAM0-vF}), we have in the Boyer-Lindquist coordinate (see equations\ (\rf{Kmetric}a,b,c,d))  
	\beeq
		\omiL=\omN+ c(\al/\vp)_{\rm iL},  \ \ \omoL=\omN- c (\al /\vp)_{\rm oL}.
		\lb{ioLS}   \eneq	
That is, there are the two (inner and outer) light surfaces, \SIL\ and \SOL, on both sides of \SSN\ where $\vF=0$ (see Figure 5 in \cite{gra14}). For an arrangement of these characteristic surfaces, the ZAMOs will see 
\beeq
\OmH\ggo\omiF>\omiL>\omN=\OmF>\omoL> \omoF \ggo 0 
\lb{GapC} \eneq  
along the same FL,  where $\omiF$ and $\omoF$ are the values of $\om$ on the inner and outer fast-magnetosonic surfaces \SoF\ and \SiF\ (see section\ \ref{Iout/in}).  
We can thus coordinatize $\OmFm$ as well as $\om$ along each FL (see, e.g.\ the horizontal axis of figure \rff{Flux-om}). 

Each FL prograde-rotates in the outer domain $\calDout$ where $\omN \ggo \om\ggo 0$, i.e., $0 \lo \OmFm \lo \OmF$, and the \emph{same} FL retrograde-rotates  in the inner domain $\calDin$ where $\OmH \ggo\om\ggo \omN $, i.e., $0\ggo \OmFm \ggo (\OmFm)_{\rm H}=-(\OmH-\OmF)$. This situation explains that in the former, the normal magnetocentrifugal wind of force-free particles flows passing outwardly through \SOL\ toward \SOF\,$\lo$\,\Sinf, while in the latter, the gravito-magneto-centrifugal wind of force-free particles flows passing inwardly through \SIL\ towards \SIF\,$\ggo$\,\SH. This inflow is never due to gravitational accretion due to the Kerr BH. 

It was already pointed out in BZ77  that ``The outer light surface corresponds to the conventional pulsar light surface and physical particles must travel radially \emph{outwards} beyond it. Within the inner light surface, whose existence can be attributed to frame-dragging and gravitational redshift, particles must travel radially \emph{inwards}.''  
It is due to the counter rotation of $\calDin$ by frame-dragging that the null surface \SSN\ must exist between the two light surfaces, \SOL\ and \SIL\, on the same FLs. Thus we see `$\om_{\rm oL} < \OmF <  \om_{\rm iL}$', which will correspond to an inequality `$\Omega_{\rm min}<\Omega< \Omega_{\rm max}$' given below equation (15) in \citet{bla22} (also see equations\ (\rf{xEM}a,b) and (\rf{xoiL}) for the behaviors of \SOL, \SSN\ and \SIL, and $\xiL\to 1$, $\xN\to 1.2599$ and $\xoL\to\infty$ for the slow-rotation limit of $h\to 0$). 

We can find a related statement in \citet{pun90} that ``the magnetospheric plasma is produced in the spacial region between \SIL\ and \SOL; at the \SIL\ (\SOL), the magnetic field rotates backwards (forwards) at the speed of light relative the plasma,'' and actually this is consistent with the existence between \SIL\ and \SOL\ of the null surface \SSN\ where $\OmFm\ggel 0$ (also see sections 7.3.1 and 9.3 in \cite{gra14}).
  
\subsubsection{The eigen-functions $\Iout$ and $\Iin$ due to the criticality condition on the fast-magnetosonic surfaces} \label{Iout/in} 

Let us determine the current/ angular-momentum function $I(\Psi)$ in the SC and GR domains. We terminate the force-free domains on the resistive membranes $\Sffinf$ and $\SffH$ near \Sinf\ and \SH, respectively (precisely speaking, on the outer and inner fast-magnetosonic surfaces \SoF\ and \SiF; see, e.g., \cite{oka78,ken83,pun90}). 
We have the behaviors of $\vcB$ and $\vcEp$ toward \Sinf\ and \SH\ by equation\ (\rf{vcBp/Bt},b) as follows; 
	\[ \vcB^2=\vcBp^2 +\vcBt^2=(2I/\vp\al c)^2+(\Bp\vp)^2/\vp^4     \] 
	\begin{eqnarray}
	\simeq  \left\{
	\begin{array} {ll} 
		(2\Iout /\vp c)^2 &; \Sffinf, \    \al \to1, \  \vp\to \infty, \\[1mm]
		(2\Iin/\vp\al c)^2 &; \SffH, \  \al\to 0, \ \om\to \OmH,
	\end{array}  \right.    \lb{Iout/inA}  \end{eqnarray} 
	\beeq 
		\vcEp^2= (\OmFm\vp/\al c)^2 \Bp^2  \nonumber \eneq 
	\begin{eqnarray}
		\simeq  \left\{
	\begin{array} {ll} 
		\left( \displaystyle{\frac{\OmF}{\vp c}}\right)^2 (\Bp\vp^2)^2 &; \Sffinf, \    \al\simeq 1, \  \vp\to \infty, \\[2mm]
		\left( \displaystyle{\frac{\OmH-\OmF}{\al\vp c}}\right)^2 (\Bp\vp^2)^2 &; \SffH, \  \al\to 0, \ \om\to \OmH,
	\end{array}  \right.    \lb{Iout/inB}  \end{eqnarray}	
where S$_{\rm oF}\lo\Sffinf$ and S$_{\rm iF}\ggo\SffH$. 
Then $(\vcB^2-\vcE^2)\to 0$ reduces to the so-called `criticality condition' 
 for $\Iout$ and $\Iin$ 
 (\cite{zna77}; MT82; \cite{oka92}), as follows; 
	\begin{subequations} \begin{eqnarray}
		 \Iout =\displaystyle{ (1/2)\OmF(\Bp\vp^2)_{{\rm ff}\infty}}  &;\ \Sffinf, \lb{Iout/in-a} \\[1mm]  
		\Iin =	(1/2) (\OmH -\OmF)(\Bp\vp^2)_{{\rm ffH}}  &;\ \SffH.    \lb{Iout/in-b}
	\end{eqnarray} \lb{Iout/in}  \end{subequations}  
The former $\Iout$ expresses the external resistance of particle acceleration on the resistive membrane $\Sffinf$ (see equation\ (\rf{I/Pul}) for $I_{\rm NS}(\Psi)$ in the pulsar case).  The latter $\Iin$ does another external resistance of entropy production on the resistive membrane $\SffH$ above the horizon, but this is not an internal resistance of a horizon battery (if any) (see figure \rff{DC-C} and section\ \ref{cir-appror}).    

By equations\ (\rf{Iout/in}a,b), the energy and angular momentum fluxes in equations\ (\rf{TotalFa,b}a,b) possess different forms in the outer and inner domains along each FL with FLAV $\OmF(\Psi)$;   
	\beeq 
	\calPEout  
	= \frac{1}{c}\int^{\bar{\Psi}}_{\Psi_0} \OmF \Iout  d\Psi,  \quad 
	 \calPEin  
	= \frac{1}{c}\int^{\bar{\Psi}}_{\Psi_0} \OmF \Iin  d\Psi 
	\lb{calPEout/in} \eneq  
	\beeq
	  \calPJout  = \frac{1}{c}\int^{\bar{\Psi}}_{\Psi_0} \Iout d\Psi, \quad 
	  \calPJin = \frac{1}{c}\int^{\bar{\Psi}}_{\Psi_0} \Iin d\Psi .
	  \lb{calPJout/in}  \eneq  
 
It must be on the null surface \SSN\ that the influx of negative angular momentum in $\calDin$ cancels out the outward flux of positive angular momentum in $\calDout$, to keep the ZAM-surface \SZAM,   
i.e., $\vcSJout= \vcSJin= -\vcSJinU$.  This condition $\Iout=\Iin$ turns out to yield the boundary condition to finally determine $\omN=\OmF$ for the whole magnetosphere frame-dragged into rotation with $\OmF=\omN$ (see section\ \ref{BC-SN}).

Next, the surface currents $\calI_{{\rm ff}\infty}$ and $\calI_{\rm ffH}$ flowing on the resistive membranes $\Sffinf$ and $\SffH$ with the surface resistivity $\RH=\calRinf=4\pi/c$ are defined as follows; 	
	\begin{subequations}  \begin{eqnarray}
		\calI_{\rm ff{\infty}}=\left(\frac{\Iout}{2\pi \vp}\right)_{\rm ff{\infty}} 
		=\left(\frac{c}{4\pi} \frac{\OmF\vp}{c} \Bp\right)_{\rm ff{\infty}} =\left(\frac{E_{\rm p}}{\calRinf}\right)_{\rm ff{\infty}},
	\lb{calIffinf}  \end{eqnarray}		
	\begin{eqnarray}  
		\calI_{\rm ffH}= \left( \frac{\Iin}{2\pi \vp} \right)_{\rm ffH} 
		=\left( \frac{c}{4\pi}  \frac{(\OmH-\OmF)\vp}{c} \Bp \right)_{\rm ffH} 
		=\left( \frac{E_{\rm p}}{\RH} \right)_{\rm ffH}.	
	\lb{calIffH} 		
	\end{eqnarray} \lb{calIinfH}  \end{subequations}  
 Ohmic dissipation of these two surface currents corresponds to particle acceleration and entropy production in each closed-circuit $\calCout$ and $\calCin$ (see equations\ (\rf{Sffinf-M},b,c)). 

\begin{figure*}
~~~~~~~~~~~~~~~~~~~~~~~~~
\includegraphics[width=11cm, height = 7cm, angle=-0]{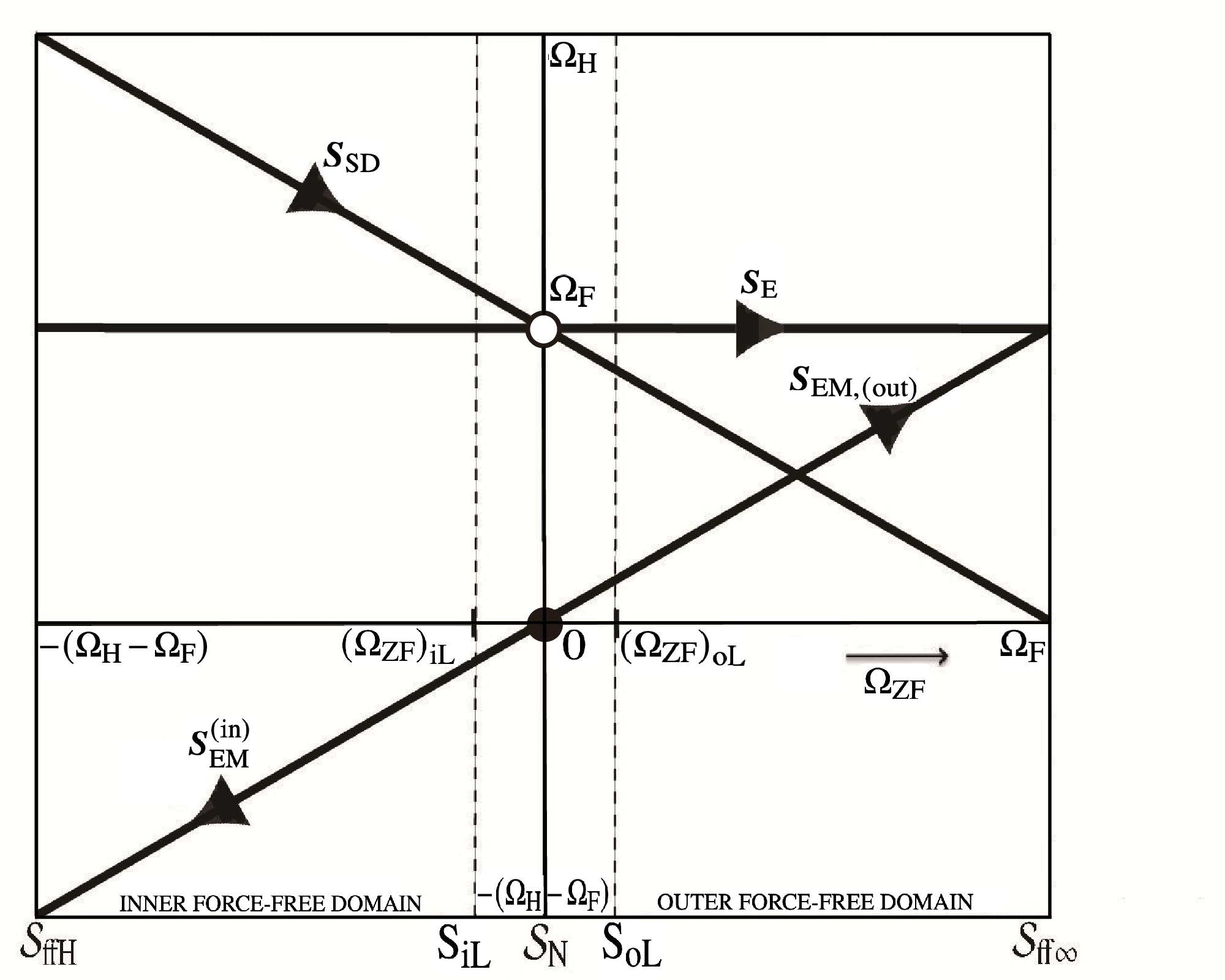} 
\caption{
The three energy fluxes $\vcSE$, $\vcSEM$ and $\vcSsd$ in equations\ (\rf{Se},b,c)) plotted against the ZAMO-FLAV $\OmFm$ along each FCSL (modified from figure 3 in \cite{oka09}). The point is to visualize the effect of the ZAMO-FDAV by coordinatizing $\OmFm$, which is a linear function of $\om$; see equation\ (\rf{GapC}).	  
The ordinate on the null surface \SSN\ ($\OmFm=0$) divides the force-free magnetosphere into the two, outer \emph{prograde}-rotating and inner \emph{retrograde}-rotating domains, $\calDout$ and $\calDin$, with the respective light surfaces \SOL\ and \SIL (see section \ref{SoL/SiL}).   Note that no reversal of the {\em conserved} flux $\vcSE=\OmF\vcSJ$ takes place, although $\vcSE=\vcSEM=\vcSsd =\vcSJ=0$ on \SSN, due to breakdown of the force-free condition (see section\ \ref{NullS} and figure \rff{GapI}). 
These energy fluxes are linked to the three terms of the 1st law of thermodynamics on the horizon under the resistive membrane $\SffH$. 	
	When the hole is accepting an influx of \emph{negative} angular momentum from the plasma-shed on \SSN\ (see section\ \ref{plasma-shed}), it looks as if the hole is launching \emph{positive} angular momentum $\vcSJ>0$ beyond \SSN\ outward to $\Sffinf$.	
 Indeed, $\vcSEM$ reverses direction because the radiation condition for the Poynting flux is given by the sign of $\OmFm\ggel 0$ for $\om\lleg \OmF$. A pair of batteries, as well as a particle source, must exist under the inductive membrane $\SN$ (see figures \rff{DC-C} and \rff{F-WS}).    }
    \lbf{Flux-om} 	\end{figure*}

\subsection{The energy flux analysis}  \label{EF-A} 

\subsubsection{Two non-conserved energy fluxes $\vcSEM$ and $\vcSsd$} \label{2Flux} 
 As seen in equation\ (\rf{twoOmFb}), the conserved FLAV $\OmF(\Psi)$ is resolved by frame-dragging to the ZAMO-FLAV $\OmFm$ and the FDAV $\om$ (both are a  `gravito-electric potential gradient', also). Substituting $\OmF$ in equation\ (\rf{twoOmFb}) into equation\ (\rf{E-AmFlux}), we have 
	\beeq
	\vcSE=\vcSEM  +\vcSsd,	 \lb{vcSE-c} 
	\eneq	  
where 
    \begin{subequations} 	\begin{eqnarray}  \ \ \ 	
    \vcSE=\OmF \vcSJ =(\OmF I/2\pi\al c) \vcBp, \lb{Se}  \\[1mm]  		
    \vcSEM=\OmFm \vcSJ=(\OmFm I/2\pi\al c)\vcBp, 	 \lb{Sem} 	\\[1mm]    
     \vcSsd =\om \vcSJ=(\om I/ 2\pi\al c) \vcBp,    \lb{Ssd}   		
     \end{eqnarray}   \lb{SE/EM/sd}  \end{subequations}	
 all of which can be reproduced from equation (4.13) in MT82 by defining the two non-conserved energy fluxes $\vcSEM=\OmFm\vcSJ$ and $\vcSsd=(\OmF-\OmFm)\vcSJ$ in there\footnote{
 The FLRV $\vF$ and the electric field $\vcEp$ in equations (5.2) and (5.3) in MT82 and also (4.32) and (4.33) in TPM86 were expressed in terms of $(\OmF-\om)$, but not in terms of $\OmFm$ (cf.\ our equations (\rf{ZAMO-Ep},c)). Also, the two non-conserved energy fluxes $\vcSEM$ and $\vcSsd$ were not given explicitly (cf.\ equation (4.13) in MT82).  
 The differences between a factor $(\OmF-\om)$ and the ZAMO-FLAV $\OmFm$ are not trivial. Unless we define $\OmFm$ explicitly, the ZAMOs will be virtually blind to perform their physical roles. 
 Relations (\rf{twoOmF}a,b) will essential not only to produce `additional physics' needed \citep{pun90}, but also to make the $3+1$-formulation (MT82) and Membrane Paradigm (TPM86) perfect toward GTED. 
 }.    
 The Poynting flux $\vcSEM$ can be simply derived from the vector product of $\vcBt$ and $\vcEp$ in equations\ (\rf{vcBp/Bt},b). Just as the overall energy flux $\vcSE$ corresponds to $c^2 dM$ in equation\ (\rf{1st-l}) for the 1st law, $\vcSEM$ and $\vcSsd$ correspond to $\Th dS$ and $\OmH dJ$ on the horizon. 

By equations\ (\rf{ZAMO-Ep}) and (\rf{Sem}) for $\vcEp$ and $\vcSEM$, we can confirm that ``Physical observers will see the electric field reverse direction on the surface $\omN=\OmF$. Inside this surface, they see a Poynting flux of energy going toward the hole. (..., when the hole is losing energy electromagnetically, this surface always exists.)'' (see the caption of Figure 2 in BZ77). Also, outside this surface ($\OmFm>0$), they will see another Poynting flux going toward infinity (see figure \rff{Flux-om}). They will then understand that ``a sufficiently strong \emph{in}flux of \emph{negative} angular momentum leaving this surface \SSN\ inwardly'' must be equivalent to ``a sufficiently strong flux of \emph{positive} angular momentum leaving the hole'' (BZ77). 
This means that the null surface \SSN\ is the zero-angular-momentum surface \SZAM\ as well (see equations\ (\rf{ZAM0-vF}) and (\rf{epsEJab}b)). 

Note that $\vcSE=\OmF\vcSJ$ is always directed outward, i.e., ``the direction of energy flow cannot reverse on any given field line unless the force-free condition breaks down" (BZ77), but the ZAMOs are now aware that 
the Poynting flux $\vcSEM$ reverses direction on the null surface \SSN, where $\OmFm\ggel 0$. Indeed, it will be argued that breakdown takes place on \SSN\ (see section\ \ref{NullS}), and a relevant pair of surface unipolar induction batteries on both sides of \SSN\ will be set up there, back-to-back and yet oppositely directed to each other. In addition,  a particle source related can be excavated under the null surface \SSN\ between the two light surfaces $\omoL$ and $\omiL$ (see sections\ \ref{SoL/SiL} and \ref{cir-appror}). 

\citet{gra14} also discussed the reversal of the Poynting flux in their spacetime approach. By taking into account the ZAMO-FLAV $\OmFm$,  
the ZAMOs will see that the surface of reversal must be the same as the surface of reversal of the particle velocity in their Figure 5, that is, the null surface \SSN, where $\OmFm=\vcSEM=\vcv=0$ (see figure \rff{Flux-om} and also Figure 2 in BZ77).

 \subsubsection{Entropy production}  \label{enttr-prod}  
 On entropy production on the resistive membrane $\SffH$, utilizing $\Iin$ in equations\ (\rf{Iout/in-b}) and $\vcSEM$ in equation\ (\rf{Sem}), we have 
	\begin{subequations}  \begin{eqnarray} 
            \TTH \frac{dS}{dt} = - \oint_{{\rm S}_{\rm ffH}} \al\vcSEM\cdot d\vcA     \ \ \lb{dS}	  \\[1mm]  
	= \frac{1}{c}\int^{\bar{\Psi}}_{\Psi_0}(\OmH-\OmF) I(\Psi) d\Psi  \\[1mm]	
		=-(\OmH-\OmFb) \dr{J}{t}>0, \hspace{2mm} \lb{ThdS/dt}    
	\end{eqnarray}  \lb{Entro} \end{subequations}      
 which leads to $\Th dS=-(\OmH-\OmFb)dJ$ in equation\ (\rf{Laws-(ii)}b). This is equal to the ohmic dissipation of the surface current $\calI_{\rm ffH}$ on the resistive membrane $\SffH$ with the resistivity $\RH=4\pi/c=377$ohm, i.e., 
	\beeq
		\TTH \frac{dS}{dt} = \int_{{\cal S}_{\rm ffH}} \RH \calI_{\rm ffH}^2 dA
		=\frac{1}{2 c} \int_{\Psi_0}^{\Psib} (\OmH-\OmF)^2(\Bp\vp^2)_{\rm ffH} d\Psi	
	\lb{H/resistance}  \eneq   
(see equation (3.99) in TPM86). The ingoing Poynting flux will not penetrate into the horizon where $I \approx 0$, with the form of a Poynting flux kept as it is. This entropy production will correspond to the amount of energy paid back to the hole as its cost of extraction. 


\subsubsection{The efficiency of energy extraction} \label{eff-EE}
The overall efficiency $\epsGTEb$ is defined by the ratio of actual energy extracted to maximum extractable energy when unit angular momentum is removed (BZ77), i.e.\ from equations\ (\rf{1st-l}), (\rf{TotalFa,b}a,b) and (\rf{GTEepsb})
	\beeq 
		\epsGTEb =\frac{(dM/dJ)}{ (\partial M/\partial J)_{S}}=\frac{\calPE}{\OmH\calPJ} =\frac{\OmFb}{\OmH}. 
	\lb{epsGTE1} \eneq  
The constraints for the energy extraction and its efficiency become by equation\ (\rf{Entro}b,c) 
	\begin{subequations} 	\begin{eqnarray}    \hspace{1cm}
		\OmH>\OmF\approx \OmFb > 0 ,  \lb{1st-(iii)} \\[1mm]      
		\epsGTE=\frac{\OmF}{\OmH} \approx \epsGTEb < 1.\lb{2nd-(iii)} 
	\end{eqnarray}     \end{subequations} 
 These inequalities ensure that ``when the hole is losing energy electromagnetically, the null surface \SSN\ on $\om=\OmF$ always exists'' (see equations (4.6) and (4.7) in BZ77). 


\begin{figure*}
\begin{center}
~~~~~~~~~~~~~~~
\includegraphics[width=12cm, height = 7cm, angle=-0]{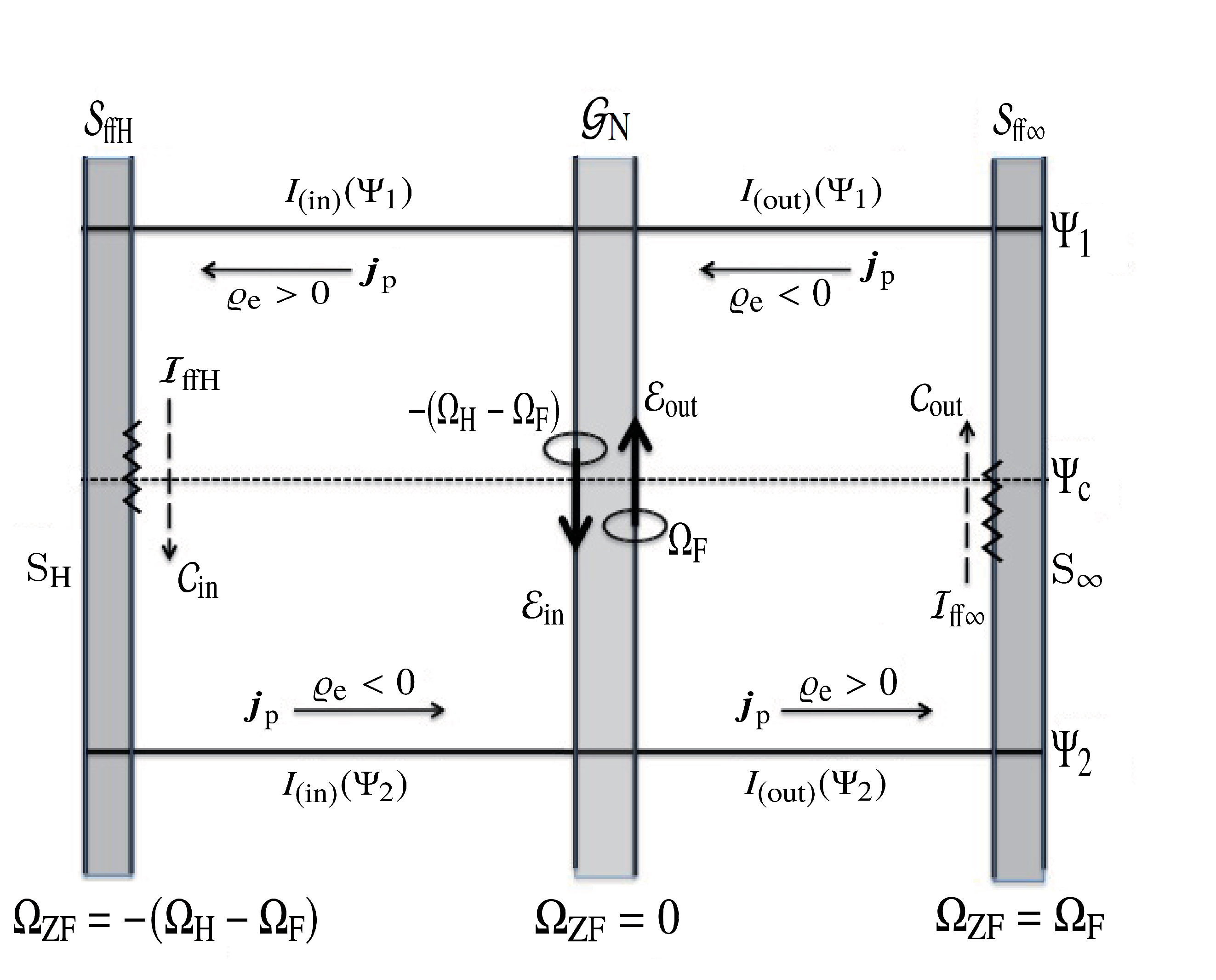}   
\end{center}
   \caption{A schematic diagram illustrating a pair of circuits $\calCout$ and $\calCin$ closed in the force-free domains $\calDout$ and  $\calDin$ \citep{oka15a}. These domains
   are separated by the null surface \SSN, where current- and stream-lines are severed by the breakdown of the force-free and freezing-in conditions (see section\ \ref{NullS}).       
   There will be dual unipolar inductors with EMFs, $\calEout$ and $\calEin$, at work related to the magnetic spin-axes oppositely directed. The AVs of these axes are $\OmF$ and $-(\OmH-\OmF)$, respectively, 
   and the difference is $\OmH$ (see  equation\ (\rf{DOmFm}); figures \rff{GapI} and \rff{F-WS}). 
   The huge voltage drop of $\Dl V\propto \OmH$ (see equation\ (\rf{DV})) will lead to viable production of charge-neutral pair-plasma towards developing a dense Gap with the half-width $\Dlom$. 
  Along the FCSL $\Psio$ where $\vcjp<0$, electrons flow out in $\calDout$, and positrons flow in $\calDin$, each charge-separated when these particles flow out from the plasma-shed to both directions (see section\ \ref{plasma-shed}). The opposite is true along the FCSL $\Psit$ where $\vcjp>0$; positrons flow outward in $\calDout$ and electrons flow inward in $\calDin$ (section\ \ref{annh}).  
   Note that $\vcvp=\vcjp/\vre>0$ in the prograde-rotating $\calDout$ and $\vcvp <0$ in the retrograde-rotating $\calDin$. 
   The particles pair-created with $\PG{\vcv}=0$, circulating around the hole's axis with $\omN$, are the ZAM-particles, dense enough to pin down magnetic field lines, to fix $\OmF=\omN$ and to make the Gap magnetized, thereby enabling the dual batteries to drive currents in each circuit. 
   It is conjectured in the twin-pulsar model (see section\ \ref{TW-P-M}) that the outer half of the Gap in $0\lo\OmFm\lo\Dlom$ plays a role of a `normal' magnetized NS spinning with $\OmF$, while the inner half in $0\ggo\OmFm\ggo -\Dlom$ behaves like an `abnormal' magnetized NS counter-spinning with $-(\OmH-\OmF)$. 
Relaxation of the RTD in-between will lead to widening of the null surface \SSN\ to the Gap $\SG$ with width $\sim 2\Dlom$ (see section\ \ref{TW-P-M}). 
}
\lbf{DC-C}  \end{figure*} 

\newcommand{\SffinfNS}{S_{\rm FF-NS}}

\subsection{The circuit analysis} 	 \label{cir-appror}   
\subsubsection{The current-closure condition}   \label{CCC}  
We impose no net gain nor loss of charges over any closed surface in the force-free domains in the steady axisymmetric state. For the closed surface from the first open FL $\Psi=\Psiz$ to the last open FL $\Psi=\bar{\Psi}$ in the poloidal plane, we have by equation\ (\rf{vcj-p})
\beeq
\oint \al\vcjp\cdot d\vcA\propto [I(\Psib)-I(\Psiz)]=0,\ {\rm i.e.},\  
I(\Psib)=I(\Psiz)=0,  
\lb{c-c-c}  \eneq	
when there is no line current at $\Psi=\Psiz$, nor $=\Psib$. This requires that function $I(\Psi)$ has at least one extremum at $\Psi=\Psic$ where $(dI/d\Psi)_{\rm c}=0$, and hence 
\beeq
\vcjp=\vre \vcvp = -\frac{1}{\al} \dr{I}{\Psi}\vcBp
\left\{
\begin{array} {ll}
<0; \ \ \Psio<\Psi< \Psic , \\[1mm]
=0;  \ \ \Psi=\Psic, \\[1mm]
>0; \ \  \Psic<\Psi<\Psit, 
\end{array}  \right. \lb{vcjpPM}  \eneq 
where $\Psiz<\Psio<\Psic<\Psit<\Psib$  (see figure \rff{DC-C} and section \ref{annh} for the two circuits $\calCout$ and $\calCin$ in the outer and inner domains). 

\subsubsection{A pair of batteries for the dual circuits}   
  
The breakdown of the two conditions (\rf{ff-a}a,b) in-between the two domains will provide an arena of setting up a pair of batteries and the voltage drop between their EMFs for particle production \citep{oka15a}. 
Let us pick up such FCSLs $\Psio$ and $\Psit$ for the circuits $\calCout$ and $\calCin$ as the two roots of an algebraic equation\  $I(\Psi)=\Iot$, i.e., $I(\Psi_1)=I(\Psi_2)\equiv \Iot$ in the range of $0<\Psio<\Psic<\Psit<\bar{\Psi}$, where $(dI/d\Psi)_{\rm c}=0$ and $\vcjp\lleg 0$ for $\Psi\lleg\Psic$. 

The Faraday path integrals of $\vcEp$ in equation\ (\rf{ZAMO-Ep}) 
along two circuits, $\calCout$ and $\calCin$, yield 
\begin{subequations} \begin{eqnarray} 
\calEout= \oint_{{\cal C}_{\rm (out)}} \al\vcEp\cdot d\vcell 
=-\frac{1}{2\pi c}\int_{\Psi_1}^{\Psi_2} \OmF(\Psi)d \Psi, \hspace{1.1cm} \lb{EMF-out} \\ 
\calEin=\oint_{{\cal C}_{\rm (in)}} \al\vcEp\cdot d\vcell
=+\frac{1}{2\pi c}\int_{\Psi_1}^{\Psi_2} (\OmH-\OmF)d \Psi. \hspace{1.1cm}  \lb{EMF-in} 
\end{eqnarray}   \lb{EMF-ab}  \end{subequations}   
There is no contribution to EMFs from integrating along $\Psio$ and $\Psit$ and on the null surface \SSN\ because of $\vcEp\cdot d\vcell=\PN{\vcEp}=0$.  
These batteries on both outer and inner surfaces of \SSN\ with no internal resistance provide electricity to the outer resistances on the resistive membranes $\Sffinf$ and $\SffH$. 
That is, the EMFs for the two DC circuits $\calCout$ and $\calCin$ drive the volume currents of charge-separated particles $\vcj$ (see equation\ (\rf{vc-vjA})) into FCSLs in the force-free domains $\calDout$ and $\calDin$ and the surface membrane currents $\calI_{\rm ff{\infty}}$ and $\calI_{\rm ffH}$ on $\Sffinf$ and $\SffH$ (see equations\ (\rf{calIinfH}a,b) and section\ \ref{annh}).  No volume current is allowed to cross the null surface \SSN\ (or the Gap $\GN$) by the breakdown of the force-free condition from one circuit to another, i.e., $\PN{\vcj}=0$ in equation\ (\rf{SNb}).

\subsubsection{Two outer and inner resistances}  
The outer resistive membrane $\Sffinf$ may also be interpreted as possessing the same surface resistivity $\calRinf=4\pi/c=377$ Ohm as on the inner membrane $\SffH$ above \SH, and Ohm's law holds on $\Sffinf$, i.e., $\calRinf \calI_{\rm ff{\infty}}=(E_{\rm p})_{\rm ff{\infty}}$. Thus similarly to equation\ (\rf{ThdS/dt}), we have by equation\ (\rf{Iout/in}b)
\begin{subequations} 	\begin{eqnarray}
\int_{ {\cal S}_{{\rm ff}\infty } } \calRinf \calI_{\rm ff{\infty}} ^2 dA 
=  \int_{ {\cal S}_{{\rm ff}\infty } }  \vccalI_{\rm ff{\infty}}\cdot\vcEp  dA  
=  \int_{ {\cal S}_{{\rm ff}\infty } } \vcSEM\cdot d\vcA  \nonumber  \\[1mm] 
=\frac{1}{2c} \int_{ {\cal S}_{{\rm ff}\infty } } \OmF^2(\Bp\vp^2)_{{\rm ff}\infty} d\Psi =\calPEout ,   \hspace{1cm} 	
\lb{Sffinf-M}  \end{eqnarray}   
where $\vcSEM=\vcSE$ and $\vcSsd=0$ on $\Sffinf$ for $\om\to 0$. On the other hand, from equation\ (\rf{ThdS/dt}), one has on the horizon resistive membrane $\SffH$
\begin{eqnarray}
\int_{{\cal S}_{\rm ffH}} \RH \calI_{\rm ffH}^2 dA= \TTH \frac{dS}{dt}=\OmH\calPJin -\calPEin. \ \ \ \  
\lb{ThdS/dt-b}  
\end{eqnarray}  	
The above two expressions sum up to
	\begin{eqnarray}
	\int_{{\cal S}_{\rm ffH}} \RH \calI_{\rm ffH}^2 dA + \int_{ {\cal S}_{{\rm ff}\infty } } \RH \calI_{\rm ff{\infty}} ^2 dA = \OmH\calPJin, \ \ \ \ 
	\end{eqnarray}	\lb{ThdS/dt-c}  \end{subequations} 
because $\calPEout=\calPEin=- c^2(dM/dt)$ and $\calPJin=\calPJout= -(dJ/dt)$ hold across the Gap $\GN$ with $\PG{I}=0$ by the boundary conditions (see equations\ (\rf{DN/SNa},b)), and hence we have $c^2 dM=\Th dS+ \OmH dJ$. It turns out thus that the first law of thermodynamics participates directly in ohmic dissipation of the \emph{surface} currents for entropy production and particle acceleration on the two resistive membranes $\SffH$ and $\Sffinf$ (see equations\ (\rf{Laws-(ii)}a,b)). 

The two EMFs in equations\ (\rf{EMF-ab}a,b) for circuits $\calCout$ and $\calCin$ are also responsible for launching the Poynting energy fluxes in both outward and inward directions, i.e., $\vcSEM\ggel 0$ for $\OmFm\ggel 0$.  Equation (\rf{EMF-out}) for $\calEout$ coincides `almost exactly' with equation\ (\rf{nsEMF}) for a pulsar magnetosphere, because of $\om\ll \omN=\OmF$ and hence $\vcSEMout\approx\vcSE$ in the outer SC domain $\calDout$. 

This ohmic dissipation implies particle acceleration. The rate per unit $\tau$ time at which electromagnetic fields transfer redshifted energy to particles is by equation (4.14) in MT82;  
	\begin{eqnarray}
		-\frac{1}{\al} \vcnb\cdot\al\vcSE=\al\vcj\cdot\vcE+(\om/c)(\vcj\times\vcB)\cdot \vcm   \hspace{0.5cm}   \nonumber \\
		=\frac{\OmF\vp}{c}\jvl\Bp  
		=- \frac{\OmF\vp}{c}\frac{\Bp}{2\pi\vp\al}\LPPlDr{\Iout}{\ell}>0. 
	\lb{DivvcSE} \end{eqnarray}  
This means that when the current function $I(\ell,\Psi)$ is continuously decreasing with $\ell$ in the resistive membrane $\Sffinf$ from near \SoF\ to \Sinf, the MHD acceleration will take place (see figure \rff{GapI}), but the force-free magnetosphere regards the `force-free' domain with $\jvl=0$ formally as extending to the force-free infinity surface $\Sffinf$ where $|\jvl|\gg |\jpl|$. By doing so, the circuit $\calCout$ closes so as to fulfil the current-closure condition in the steady state. 

\subsubsection{Particle production due to voltage drop}  \label{PP-V} 

The difference between the two EMFs in equation\ (\rf{EMF-ab}a,b) across \SSN\ is 
	\beeq
		\DN{\calE}=\calEout-\calEin=-(\OmH \Dl \Psi/2\pi c)=-\Dl V,
		\lb{DV}  \eneq   
where $\Dl\Psi=\Psio-\Psit$, and the difference of a quantity $X$ across \SSN\ is denoted with
	\beeq	\DN{X}=(X)_{{\rm N}{(\rm out)}}  - (X)_{{\rm N}{(\rm in)}}. 
	 \lb{DfN}  \eneq 
The difference of the ZAMO-FLAV $\OmFm$ between \Sinf\ and \SH\ becomes from equation\ (\rf{twoOmFa}) 
 \beeq 
 (\OmFm)_{\infty}-(\OmFm)_{\rm H} 
 =\OmF- [-(\OmH-\OmF)]=\OmF + (\OmH-\OmF)=\OmH.
  \lb{DOmFm}  \eneq 

The `spark' models so far used for pair-production discharges in literature are based mainly on an extension from a `negligible violation' of the force-free condition (BZ77; MT82; \cite{phi83a,bes92,hir98,son17,hir18,sit24}; see also e.g.\ \cite{ruf10,che23} for a general review). 

It is argued here that the `complete' violation of the force-free condition due to frame-dragging on the null surface \SSN\ leads to a unique gap model for the particle-current sources. It was emphasized in 2015 that ``the present gap model with a pair of batteries and a strong voltage drop fundamentally differs from any existing models based on pulsar outer-gap models.'' \citep{oka15a}.  The significant differences from the previous particle production mechanism come mainly from the existence of the counter-rotating inner domain $\calDin$ due to frame-dragging, with $\epsJ\leq 0$ inside \SSN\ (see section\ \ref{TW-P-M}). 

The null surface \SSN\ dividing the force-free magnetosphere into the two (GR and SC) domains seems to be genetically endowed with the discontinuity $\Dl(\OmFm)_{\rm N}\approx (\OmFm)_{\infty}-(\OmFm)_{\rm H}=\OmH$ and $\DN{\calE}=-\Dl V$, to widen the surface \SSN\ to a gap $\SG$ under $\SN$, thereby constructing a magnetized `\zam' and `charge-neutral' Gap between the two force-free domains with $\epsJ\ggel 0$ (see section\ \ref{m-mdGap}). 
Thus, the voltage drop $\Dl V$ in equation\ (\rf{DV}) reveals that the null surface \SSN\ will be a kind of rotational-tangential-discontinuity (RTD) due to the two (inner and outer) magnetic rotators, namely between the two, counter-rotating each other, outer and inner domains $\calDout$ and $\calDin$,  although $\OmFm$ and $\vcEp$ seem to change sign smoothly through zero (see section\ \ref{TW-P-M}; cf.\ \cite{lan84}). It turns out thus that the maximum available voltage drop $\Dl V=(\OmH/2 \pi c)\Dl\Psi$ will be utilizable in-between the two circuits $\calC_{\rm out}$ and $\calC_{\rm in}$ in the steady state (see section\ \ref{TW-P-M}). 

 When the ZAM-Gap $\GN$ is charge-neutral, i.e., $\varrho_{\rm e} \approx 0$ as a result of ample pair-creation in the steady state, it will be another role of a pair of batteries that drive charge-separated particles from pair-created, charge-mixed plasma into each FCSL in the force-free domains $\calDout$ and $\calDin$ (see figures \rff{DC-C} and \rff{GapI}; section\ \ref{annh}).

\section{The energy and angular-momentum densities of the electromagnetic fields} \label{EnegR}	 
\setcounter{equation}{0}
\numberwithin{equation}{section}  

For the densities of the field energy and angular momentum, substituting $\vcB$ and $\vcEp$ from equations\ (\rf{vcBp/Bt},b) into equations (2.30a) and (2.31a) in MT82, we have
  \begin{subequations} \begin{eqnarray}
	\epsE= \frac{\al\Bp^2}{8\pi} \left[1+\frac{\Bt^2}{\Bp^2} +\frac{\vp^2}{\al^2 c^2}(\OmF^2 -\om^2 )\right] ,  \lb{epsEba} \\ 
	\epsJ=\frac{\vp\vF}{c}\Bp^2 = \frac{\OmFm (\vp\Bp)^2}{\al c}  \lb{epsJbb}  
	\end{eqnarray} \lb{epsEJab}  \end{subequations}  
(also see equation (2.17a) in \cite{oka92} and equation (55) in \cite{kom09}), where $\epsE$ and $\epsJ$ are regarded as an `explicit' function of $\om$ along each field FL labeled with $\Psi$. The angular momentum density $\epsJ$ changes sign on the null surface \SSN, and we refer to the zero-angular-momentum-density surface as \SZAM, which accords with \SSN. 

Near the null surface \SSN\ where $\om=\OmF$ and $\vp\Bt\propto I= 0$ (see section\ \ref{NullS}), we see 
        \beeq
        \epsE=\left(\frac{\al \Bp^2}{8\pi}\right)_{\rm N}>0, \ \ \epsJ = 0.
        \lb{epsE/N} \eneq  
The field energy will be strong enough to magnetize the plasma pair-produced with the voltage drop $\DN{\calE}=-\Dl V$ between the two EMFs in equation\ (\rf{DV}). Conversely, the plasma density will be large enough to keep the field $\vcBp$ frozen-in to ensure the magnetosphere frame-dragged by $\OmF=\omN$ (see sections\ \ref{m-mdGap} and \ref{BC-SN}). 

	It is the frame-dragging term $\om^2$ in equation\ (\rf{epsEba}) that builds a negative-energy region with $\epsE<0$ in the inner domain $\calDin$. 
At the inner light surface \SIL, where $\vF=-c$ and $\omiL$ is given by equation\ (\rf{ioLS}). When $(\Bp\vp^2)_{\rm ffH}/ (\Bp\vp^2)_{\rm iL}\approx 1$ and $\OmF\approx 0.5\OmH$, we have analytically from equation\ (\rf{epsEba}) 
    \beeq 
     \epsE \approx -\left( \frac{\Bp^2}{8\pi}\frac{\OmF\vp}{c} \right)_{\rm iL} 
        \left[2- \left(\frac{\OmF\vp}{c\al} \right)_{\rm iL} \right],  
		\eneq 
and then we see $(\epsE)_{\rm iL}<0$, if $(\OmF\vp/c\al)_{\rm iL}<2$, i.e., $\omN<\omiL/2$. 
	
Also we see in equation\ (\rf{epsEba}) that there will be such a surface $\SepsE$ that divides the inner domain $\calDin$ farther into the two regions by $\epsE(\om,\Psi)\ggel 0$ for $\om \lleg \omepsE$ between \SSN\ and $\SffH\approx$\SH, where $\omepsE$ is a solution of equation for $\epsE=0$ in equation\ (\rf{epsEba}), i.e., 
\beeq
\omepsE^2 = \omN^2 +\frac{\al^2 c^2}{\vp^2} \left(1+\frac{\Bt^2}{\Bp^2} \right),  
\lb{NER} \eneq 
where $\al/\vp$ and $\Bt/\Bp$ are thought of as functions of $\om$ and $\Psi$. 
Therefore, it is the frame-dragging that produces not only the inner domain $\calDin$ of $\OmFm \leq 0$ with \SIL but also a region of the negative-energy density of $\epsE\leq 0$ in $\OmH\geq\om\geq \omepsE$. We guess $\omepsE\approx \omiL$. 	
	
 Near the resistive horizon membrane $\SffH$ where $(\Bp\vp^2)_{\rm ffH}/(\Bp\vp^2)$ $\approx 1$ and hence $\Bt^2/\Bp^2\approx 
 (2\Iin/\vp c\al\Bp)^2\approx ((\OmH-\OmF)\vp/(\al c))^2$ by equation\ (\rf{Iout/in-b}), we have  $$ \epsE \approx -\left(\frac{\Bp^2}{4\pi \al} \frac{\OmF\OmH\vp^2}{c^2}\right)_{\rm ffH} \left[\left(1-\frac{\OmF}{\OmH}\right) -\left( \frac{c^2\al^2}{2\OmF\OmH\vp^2}\right)_{\rm ffH} \right] $$ 
	\beeq
	\approx - \left(1-\frac{\OmF}{\OmH}\right)\left(\frac{\Bp^2}{4\pi \al} \frac{\OmF\OmH\vp^2}{c^2} \right)_{\rm ffH} <0   
	\lb{epsE/ffH} \eneq  
for $\al\to 0$ toward the resistive horizon membrane $\SffH$.

For the density of angular momentum near the horizon, we have
	\beeq
	\epsJ=-(\OmH-\OmF) \LPfrac{ (\vp\Bp)^2}{\al c}_{\rm ffH}<0.   
	\lb{epsJffH}  \eneq  
It will be evident in equation\ (\rf{NER}) that $\OmH> \omepsE >\omN=\OmF$, which does not lead to any more robust condition upon $\OmF$ than that from the second law and the radiation condition toward the horizon. 

The above result suggests that the negative-energy region will indeed extend from near \SIL\ to \SH\ in the inner domain $\calDin$. 
However, the existence of the negative energy region near the horizon will not be an exact or direct indicator of energy extraction taking place from a Kerr hole. Noteworthy is the evidence that the force-free magnetosphere is divided into the two (prograde-rotating SC and retrograde-rotating GR) domains by the null surface \SSN, i.e., the Zero-Angular-Momentum-Density surface \SZAM\ where $\epsJ=\vF=\OmFm=0$ and $\epsE>0$ (see equations\ (\rf{epsE/N})). 

\section{Breakdown of the force-free and freezing-in conditions} \label{NullS}  
\setcounter{equation}{0}
\numberwithin{equation}{section} 
The force-free condition must be broken down somewhere in the active force-free magnetosphere of a Kerr BH \citep{uch97a}.
The ZAMOs will see that the electric field $\vcEp$ and hence the Poynting flux $\vcSEM=\OmFm\vcSJ$ reverse direction on every FLs threading the null surface \SSN, because of the counter-rotation of the inner GR domain ($\OmFm<0$) against the outer SC domain ($\OmFm>0$). Also, they will understand why a pair of unipolar-induction batteries $\calEout$ and $\calEin$ must be set up on the outer and inner sides of the null surface \SSN, oppositely directed (see figure \rff{DC-C}). 
These pieces of evidence obviously require the particle source as well as the current source under the null surface \SSN\ with $\omN=\OmF$  and $\vcEp=\OmFm=0$
\footnote{We can confirm the indication of showing for the vector $\vec{E}$'s reversal of direction in Figure 2 and its caption in BZ77, and in Fig.\ 38 in TPM86, as well as in each ordinate of figures \rff{Flux-om}, \rff{DC-C}, \rff{GapI}, \rff{F-WS} on the null surface \SSN, where $\vcEp=\OmFm=0$ and $\vcj=\vcv=0$ (see also Figure 5 in \cite{gra14}). }.   

Then, when FLs thread this surface \SSN, i.e., $\PN{\vcBp}\neq 0$ and $\PN{\OmF}\neq 0$, the following quantities must necessarily vanish on \SSN; 
      \begin{subequations} \begin{eqnarray} 
		\PN{\vcEp} =\PN{\vcSEM}=\PN{\vre}=\PN{\vF}=\PN{\epsJ} =0,  \hspace{0.3cm} \ \ \ \ \ \ \ \lb{SNa} \\  
		\PN{\vcj}=\PN{I}  =\PN{\Bt} =\PN{\vcSJ} =\PN{\vcSsd}=\PN{\vcSE}=0, \hspace{0.3cm} \ \ \ \ \ \ \ \lb{SNb} \\  
		\PN{\calPE}=\PN{\calPJ}=0,  \hspace{1cm}  	\lb{SNd} \\  	
		\PN{\vcv}=\PN{\vcj/\vre}=0,  \hspace{1cm}  	\lb{SNc}   	
		\end{eqnarray} \lb{EqSN} \end{subequations}    
where equations\ (\rf{ff-a}a,b), (\rf{vcj-rhoe}a,b,c), (\rf{vc-vjA}), (\rf{Se},b,c) and (\rf{epsJbb}) are used, We denote the value of any function $X(\OmFm,\Psi)$ on \SSN\ where $\OmFm=0$ as follows;   
\beeq
\PN{X}=X(0,\Psi). 
\lb{PNX} \eneq   


The above Constraints unequivocally require us to rebuild the whole force-free magnetosphere from a `single-pulsar model' to a `twin-pulsar model', as follows:  
\benu   
\item  
The most important of the above Constraints will be $\PN{\vcj}=\PN{\vcv}=0$, which require current- and stream-lines to be severed on the null surface \SSN. That is, the current-wind system is separated into the two (outer SC and inner GR) domains by the null surface \SSN, i.e., the magneto-centrifugal divider with $\epsJ=\vF=\OmFm=0$.  It is to accommodate the particle-current sources on the null surface \SSN\ by breaking down the force-free and freezing-in conditions not negligibly (BZ77; MT82; TPM86) but completely. 

\item 
Constraint $\PN{\vcj }=0$ plays the role of a perfect circuit breaker as a safety device to block acausal currents across \SSN\ from a horizon battery (if any) to external resistances such as particle acceleration. This instead indicates the necessity of a pair of `surface batteries' back-to-back at both sides of \SSN, yet oppositely directed, with the particle source in-between, in which the voltage drop between the two EMFs will produce pair-particles (see section\ \ref{m-mdGap} and figure \rff{DC-C}). 

\item 
Constraint $\PN{\vcv}=0$ means that the particles pair-created under \SSN\ with $\epsJ=0$ are `\zam' particles circulating with $\omN=\OmF$. Then, Constraints $\PN{I}=\PN{\vcSJ}=\PN{\vcSE}=0$ mean that no angular momentum nor energy is transported across \SSN, even when the FLs are continuous. It is helpful to remind that the toroidal field $\Bt$ is a swept-back component of the poloidal component $\vcBp$ due to inertial loading in the resistive membranes $\Sffinf$ and $\SffH$ (see figure \rff{GapI}).  
$\PN{I}=0$ means that there must be a jump of $I(\Psi)$ from $\Iin$ to $\Iout$, just like in the NS surface (see equations\ (\rf{I/Pul-M}) and (\rf{OL-I})). Thus, we may impose the boundary condition $\DG{I}=0$ (see equation\ (\rf{DN/SN})).  

\item 
Although the current $\vcj$ does not reverse direction, the velocity $\vcv$ does, i.e., $\vcv\ggel 0$ and $\epsJ\ggel 0$ in equation\ (\rf{epsJbb}) for $\OmFm\ggel 0$ (see section\ \ref{plasma-shed}, figure \rff{F-WS}). 
Thus, the surface \SSN\ will behave like a watershed in a mountain pass (i.e.\ `plasma-shed') between outflows and inflows of `force-free' and `charge-separated' particles pair-created by the voltage drop (see equations\ (\rf{EMF-ab}a,b) and (\rf{DV})), and yet both flows are due to the magneto-centrifugal forces at work toward the opposite directions, inward and outward by $\OmF\ggel 0$, respectively. 
As the outer pulsar-type magneto-centrifugal wind flows through \SOL\ in $\calDout$ with $\vF> 0$, the inner anti-pulsar-type wind will pass through \SIL\ in $\calDin$ with $\vF< 0$. 

\item  
Since $\vre\vcEp$ vanishes but does not change the sign on \SSN, this reacts back on the force-free condition in equation\ (\rf{ff-fz}), producing just $\PN{\vcj }=0$. In contrast, the change in the direction of $\vcEp\propto\OmFm$ across \SSN\ is taken over the particle velocity $\vcv$ as it is, i.e., $\PN{\vcv}\ggel 0$ for $\OmFm\ggel 0$ in the freezing-in condition (\rf{fi-c}). This is because axial symmetry $\vcE_{\rm t}=0$ will lead to $\vcvp=\kappa \vcBp$ and  $\kappa=-(1/\vre \al)(dI/d\Psi)=0$ on \SSN\ (see equation\ (\rf{kappa})), and hence the ZAMOs will see that the particle velocity $\vcv$ behaves like $\OmFm\ggel 0$ across \SSN, contrary to $\vcjp$. Note that when $\vcEp\propto \OmFm$ across \SSN, this nature must straightly be succeeded to the particle velocity, i.e., $\vcvp\propto \OmFm$ as well.  

\item 
There will be no single circuit allowed, with such a current as crossing \SSN\ due to a single battery at any plausible position. Each electric circuit must close within its respective force-free domain ($\calDout$ or $\calDin$), with each EMF ($\calEout$ or $\calEin$) in equation\ (\rf{EMF-ab}a,b), and with each eigenvalue $I(\Psi)$ ($\Iout$ or $\Iin$) in equation\ (\rf{Iout/in}a,b). 

\item 
Two vectorial quantities $\vcjp$ and $\vcSJ$ are closely related to each other through the current/angular-momentum function $I(\Psi)$, i.e., two-sidedness in the force-free domains. It seems that both do not reverse direction, despite that $\PN{\vcjp}=\PN{\vcSJ}=0$. This is because an outflow of negative charges means the ingoing current, and an inflow of negative angular momentum implies an outflow of positive one (see figures \rff{DC-C} and  \rff{GapI}). Despite that, the null surface \SSN\ exists always. Yet, Constraints $\PN{\vcSJ}=\PN{\vcSE}=0$ hold, the overall energy flow $\vcSE=\OmF\vcSJ$ seems to flow outwards all the way along each open FL, apparently as if crossing \SSN, where the force-free condition `breaks down' on \SSN, i.e., $\PN{I}=0$, and indeed the Poynting flux $\vcSEM$ reverses direction. These are pretty natural tactics stemming from the supreme order of keeping the pair-producing Gap in the zero-angular-momentum and charge-neutral state. 
\enen

\newcommand{\vreN}{(\vre)_{\rm N}}
\begin{figure*}
\begin{center}
\includegraphics[width=12cm, height = 7cm, angle=-0]{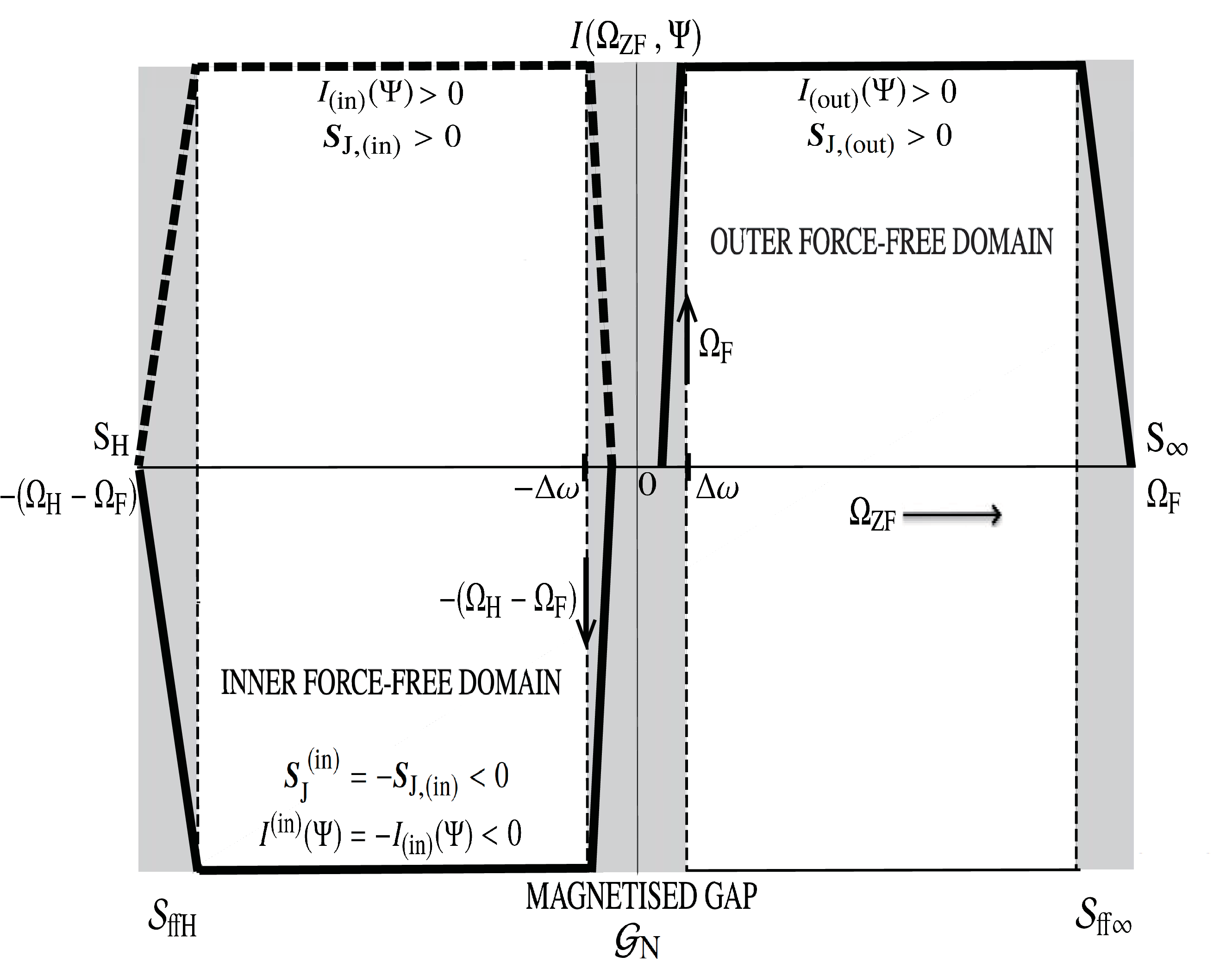}  
\end{center}
\caption{
A plausible behavior of the angular-momentum-flux/current function, $I(\OmFm,\Psi)$ (see equation\ (\rf{OL-I})). 
The force-free magnetosphere will be divided by the inductive membrane $\SN$ into the two domains of pro- and retro-grade rotation with $I=\Iout(\Psi)$ and $=\Iin(\Psi)$ in equations\ (\rf{Iout/in}a,b). 
The voltage drop $\Dl V$ between the two EMFs in equation\ (\rf{DV}) will produce pair-particles, thereby developing a Gap $\GN$ with $\PG{I}=0$ in a finite zone of no internal resistance $|\OmFm|\lo \Dlom$ between the two domains. 
Current- and stream-lines are no longer allowed to thread $\SG$ by 
Constraints 
$\PG{\vcv}=\PG{\vcj}=0$ in equations\ (\rf{SGa}).   
When the rate of \emph{positive} angular momentum conveyed outwardly into $\calDout$is equal to that of \emph{negative} angular momentum conveyed inwardly into $\calDin$, the `ZAM-state' of the Gap $\PG{\epsJ}=0$ will be maintained, i.e., $\DG{I}=\Iout-\Iin=0$. The `boundary condition' $\DG{I}=0$ in equation\ (\rf{DN/SN}) yields the eigenfunction $\OmF(\Psi)=\omN$ in equation\ (\rf{EigenOmFI}). 
The Gap filled with ZAM-particles will ensure the pinning-down of poloidal field lines $\vcBp$ with $\OmF=\omN$, and the pinning-down conversely ensure magnetization of ZAM-particles with $\PG{I}=0$ inside the Gap (see section\ \ref{PPP}). 
}
\lbf{GapI}  \end{figure*}    

\section{The zero-angular-momentum and charge-neutral gap $\GN$} \label{m-mdGap}  
\setcounter{equation}{0}
\numberwithin{equation}{section}
The ZAM-gap $\GN$ under the inductive membrane $\SN$ will play a hub of the Kerr hole's magnetospheric activities. 
 
\subsection{A plausible gap structure}  \label{GapStruc} 
We presume that the RTD with the voltage drops $\DN{\calE}=-\Dl V$ in the force-free limit will be relaxed as a result of pair-particle production to a ZAM-Gap $\GN$ with a half-width $\Dlom$. For the `widened' null Gap $\SG$ in $|\OmFm|\lo\Dlom$, we replace Constraints in equation\ (\rf{EqSN}) on \SSN\ as they are;
\begin{subequations} \begin{eqnarray} 
\PG{\OmFm}= \PG{\vcEp}=\PG{\vre} = \PG{\vcj}=\PG{\vcv}= \PG{I} \hspace{0.6cm}   \lb{SGa} \\ 
=\PG{\epsJ}=\PG{\vcSJ}=\PG{\vcSEM}=\PG{\vcSsd}=\PG{\vcSE} \hspace{0.6cm}   \lb{SGb} \\
=  \PG{\calPE}= \PG{\calPJ} =0.	 \hspace{0.6cm}   \lb{SGc} 
\end{eqnarray} \lb{EqSG} \end{subequations}    
As before, $\DG{X}$ denotes the difference of $X(\OmFm, \Psi)$ across the Gap width $2\Dlom$ (cf.\ $\DN{X}$ in equation\ (\rf{DfN}));  
\beeq 
\DG{X} =X(\Dlom,\Psi)-X(-\Dlom,\Psi) \equiv (X)_{{\rm G}{(\rm out)}}  - (X)_{{\rm G}{(\rm in)}} .    
\lb{DGX}   \eneq   
We think of such a simple form of $I=I(\OmFm,\Psi)$ along a typical open FL in $\Psiz\leq\Psi\leq\Psib$ as follows; 
 \begin{eqnarray}
 I(\OmFm) =  \left\{
\begin{array} {ll}
\to 0 & ;\ \Sffinf\ ( \OmFm\to\OmF), \\ [1mm]
\Iout & ;\ \calDout\ ( \Dlom \lo \OmFm \lo \OmF),  \\[1mm]   
0 & ;\ \SG\ ( |\OmFm|\lo \Dlom) , \\[1mm] 
\Iin &;\ \calDin\ (-\Dlom \ggo \OmFm\ggo-(\OmH-\OmF)),  \\[1mm]
\to 0 & ;\ \SffH\  (\OmFm\to -(\OmH-\OmF)) 
\end{array}  \right.    \lb{OL-I}  \end{eqnarray}    
(see figure \rff{GapI}), where $\Iout$ and $\Iin$ are given by equations\ (\rf{Iout/in}a,b).  The behaviour of $I(\OmFm,\Psi)$ in the outer domain $\calDout$ will be similar to that of the force-free pulsar magnetosphere (see equation\ (\rf{I/Pul-M})). When the Gap $\GN$ of the particle source will be situated well inside the light surfaces, we have by equation\ (\rf{ioLS})
\beeq
|\OmFm| \lo \Dlom \ll c(\al/\vp)_{\rm oL} \approx c(\al/\vp)_{\rm iL}. 
\lb{SG<SoL} \eneq   
It is not clear now how helpful or rather indispensable is the above condition in constructing a reasonable gap model, but we presume that the particle production will eventually take place intensely by the voltage drop across the Gap $\GN$, $\Dl V=-(\OmH/2\pi c)\Dl\Psi$, almost independently of the presence of the two light surfaces in wind theory. 

The Gap $\GN$ under the inductive membrane $\SN$ must be in the ZAM-state $\PG{\epsJ}=0$, so that the particles and the field carry no angular momentum nor energy across the Gap, i.e., $\PG{\vcSJ}=\PG{\vcSE}=0$ in equation\ (\rf{SGb}), when the Gap is threaded by poloidal field lines, i.e., $\PG{\vcBp}\neq 0$.  Also, the Gap must be magnetized (i.e., $\PN{\epsE}\neq 0$; see equation\ (\rf{epsE/N})), in almost the same sense as a magnetized NS because the poloidal magnetic field lines (with no toroidal component) threading the Gap will naturally be pinned down in the ZAM-particles pair-created and circulating round the hole with $\omN=\OmF$ (see section\ \ref{PPP}).  Therefore, the magnetized ZAM-particles will ensure $\OmF=\omN$ (see section\ \ref{BC-SN}). The non-force-free magnetized ZAM-Gap $\SG$ where $\PG{\vcj}=\PG{\vcv}=0$ but $\PG{\vcBp}\neq 0$ will therefore be formed in the steady-state, with its `surfaces' \SgapO\ and \SgapI, respectively, at $\OmFm\approx \pm\Dlom$ (see figure \rff{GapI}), where $\Dlom\approx|\PN{\partial\om/\partial\ell}| \Dlell$ stands for the Gap half-width (see equation\ (\rf{OL-I})), and for $\Dlom\to 0$, $\SG\to$\SSN\ (see figure 4 in \cite{oka15a} for the interplay of microphysics with macrophysics in the magnetized, matter-dominated Gap). Thus we conjecture that the voltage drop, $\Dl V=-\DN{\calE}$ on \SSN\ in equation\ (\rf{DV}), will produce pair particles copious enough, and the plasma pressure in the steady-state will expand \SSN\ to $\GN$ with a half-width $\Dlom$ in $|\OmFm|\lo\Dlom$. 

\subsection{Pinning-down of threading field lines on ZAM-particles and magnetization of the matter-dominated Gap}  \label{PPP} 
When we regard the Gap surfaces \SgapO\ and \SgapI\ as being equipped with EMFs $\calEout$ and $\calEin$, respectively, these EMFs will not only drive currents in the respective circuits $\calCout$ and $\calCin$ but also produce a strong voltage drop $\Dl V=-\DG{\calE}$ across the Gap, which will create copious ZAM-particles necessary to pin threading FLs down on. The ZAM-Gap filled with ZAM-particles will then circulate the hole with $\omN=\OmF$, and the poloidal field lines threading the Gap $\SG$ will indeed be pinned down on ZAM-particles with $\OmF=\omN$. Thus, the ZAM-Gap will be in the perfectly magnetized state, with no electric current and no angular momentum flux allowed to cross, i.e., $\PG{\Bt}=\PG{I}=\PG{\vcj}=0$. The physical state of the ZAM-Gap $\GN$ will be analogous to that of the NS inside, ensuring the boundary condition $\OmF=\OmNS$ for the FLs emanating from the NS surface. 

  
\subsection{Magneto-centrifugal plasma-shed on the ZAM-surface}  \label{plasma-shed} 
The ZAMOs circulating with $\om$ will see the force-free magnetosphere as follows: 
The outer domain $\calDout$ behaves like a pulsar-type magnetosphere rotating with $\OmF$, whereas the inner domain $\calDin$ will act like an anti-pulsar-type magnetosphere rotating with $-(\OmH-\OmF)$. Then, plasma particles pair-created in-between by the voltage drop $\Dl V=- \DG{\calE}$ circulate at $\om=\OmF=\omN$ and may not behave as force-free particles with negligible inertia within the Gap. These ZAM-particles with $\PG{\vcv}=0$ will soon become charge-separated inside two batteries to flow from the Gap out to the two force-free domains as electric charges along current lines as well as wind particles along stream-lines (see figures \rff{DC-C} and \rff{F-WS}). 
The null surface \SSN$=$\SZAM\ midst the ZAM-Gap $\GN$ redefines quite a new general-relativistic type of divider due to magneto-centrifugal force modified by frame-dragging for particles pair-created in the spark ZAM-Gap, outward and inward ($\vcv\ggel 0$).  That is, this surface \SZAM\ will play the role of a magneto-centrifugal plasma-shed, akin to a gravitational water-shed of a mountain ridge for heavy rainfalls on the Earth.  
 This will be quite a natural way to launch `magneto-centrifugal' winds from the ZAM-Gap for \emph{both} directions toward infinity ($\vcv>0$) and the horizon ($\vcv<0$) (see section\ \ref{Wind-A}), similarly to the Poynting flux ($\vcSEMout>0$) for particle acceleration on the resistive membrane $\Sffinf$ and the one ($\vcSEMin<0$) for entropy production on another membrane $\SffH$ (see section\ \ref{EF-A}).

\subsection{Pair-creation, charge-separation and pair-annihilation } \label{annh}
One of the important properties of the `force-free' plasma will be `charge-separatedness'. We will be able to utilize this in another way of discharging a pair of batteries on the null surface \SSN\ into the two resistive membranes $\Sffinf$ and $\SffH$, as something like an electron-positron 
collider. 

Let us consider that strong enough pair-creation due to the voltage drop in the spark gap between the two EMFs will be at work in producing copious pair-particles (e.g., $\gamma+\gamma\to e^{-}+e^{+}$).  These mixed charges of $e^{\pm}$ in the Gap $\GN$ will then be charge-separated in the presence of the battery EMF $\calEout$ in the outer circuit $\calCout$, as the $e^{-}$-stream flowing from under \SSN\ outward into the `FCS-line $\Psio$', and similarly as the $e^{+}$-stream flowing from under \SSN\ outward into the `FCS-line' $\Psit$ (see the arrows of $\vcjp$ in figure \rff{DC-C}). This is because the wind always blows outward in the outer domain ($\OmFm>0$ and $\vcv>0$). Also, by the effect of the battery $\calEin$, $e^{+}$- and $e^{-}$-streams from \SSN\ will flow inward into the same FCSLs, $\Psio$ and $\Psit$, respectively, in the inner circuit $\calCin$ ($\OmFm<0$ and $\vcv<0$). 

 These two $e^{\pm}$-streams in the two circuits $\calCout$ and $\calCin$ will collide with each other in the restive membranes, $\Sffinf$ and $\SffH$,  respectively, and then pair-annihilation will take place, thereby liberating energy of the same order of magnitude as consumed when pair-particles are created in the Gap (i.e., $e^{-}+e^{+}\to \gamma+\gamma$).  This will be comparable to the amount of energy due to ohmic dissipation of the surface 
 current $\calI_{{\rm ff}\infty}$ 
 and $\calI_{\rm ffH}$ in order of magnitude (see equations\ (\rf{Sffinf-M},b,c)). Also, the sum of the energy liberated in $\Sffinf$ and $\SffH$ will be similar to the amount due to the sum of the Poynting flux $\vcSEM$, inward and outward (see equation\ (\rf{SDenergy}) later).
 

 

\section{The eigen-magnetosphere}  \label{BC-SN}  
\setcounter{equation}{0}
\numberwithin{equation}{section}

For a viable force-free magnetosphere, we refer to the condition by which to finally determine the eigenfunction $\OmF(\Psi)=\omN$ as the boundary condition, distinguishing from the criticality condition for another eigenfunction $I(\Psi)$ in equations\ (\rf{Iout/in}a,b). 

\subsection{The boundary condition for the eigenfunction $\OmF$} \label{BCagain}   
One of the vital roles of the ZAM-Gap is to anchor the poloidal field $\vcBp$ onto the ZAM-particles pair-created in there and to accomplish magnetization of the ZAM-Gap, thereby ensuring $\OmF=\omN$ for threading field lines. Accordingly, the ZAM-state of the Gap must always be maintained in the magnetosphere frame-dragged by the hole into circulation with $\omN=\OmF$. 

We formulate the `boundary condition' with Constraints (\rf{EqSN}) or (\rf{EqSG}) and  with postulate (\rf{OL-I}) appropriately taken into account, that is, $\PG{I}=\DG{I}=0$ at the place of the ZAM-Gap ; 
	\begin{subequations} \begin{eqnarray}
		\DG{I}=\Iout(\Psi) -\Iin(\Psi) \quad \quad  \lb{DN/SNa} \\   
		=\Iout(\Psi) +\IinU(\Psi)=0  \lb{DN/SNb}.       
	\end{eqnarray}  \lb{DN/SN}  \end{subequations}     
This ensures the continuity of overall energy flux as well as angular momentum flow across the Gap, thereby keeping the charge-neutral ZAM-state  
along each FL threading the Gap (see figure \rff{GapI}; equations\ (\rf{Iout/in}a,b)). 

Condition (\rf{DN/SNa}) shows that the outward transport rate of \emph{positive} angular momentum leaving \SgapO\ into the SC domain $\calDout$ must be equal to that entering \SgapI\ from the GR domain $\calDin$. This is because the energy and angular momentum flows do not take place actually inside the ZAM-Gap with $\PG{I}=0$. Condition (\rf{DN/SNb}) implies equivalently that the outward rate of {\em positive} angular momentum leaving \SgapO\ is offset by the inward rate of {\em negative} angular momentum leaving \SgapI\ toward the hole. 

Now, by equations\ (\rf{E-AmFlux}), we have
	\beeq
		\DG{\vcSE}=\OmF\DG{\vcSJ}=0,
	\lb{DG/SE/SJ}  \eneq  
which apparently shows that the overall energy and angular momentum fluxes flow outward continuously across the Gap $\GN$, regardless of $\PG{\vcSE}=\PG{\vcSJ}=0$. Likewise, the `boundary condition' (\rf{DN/SN}a,b) ensures no discontinuity of the power $\calPE$ and the loss rate of angular momentum $\calPJ$ across the ZAM-Gap, i.e., as shown by equations\ (\rf{TotalFa,b}a,b) and (\rf{calPEout/in}) and (\rf{calPJout/in});
	\begin{subequations}  	\begin{eqnarray}
		\DG{\calPE}=\calPEout- \calPEin  =\calPEout + \calPEinU =0, \hspace{1cm}  \lb{DGcalE}   \\   
		\DG{\calPJ}=\calPJout -\calPJin  =\calPJout + \calPJinU =0. \hspace{1cm}  \lb{DN/calPJ} 
	\end{eqnarray}  \lb{DN/calPEJ}  \end{subequations}     

Analysis of wave propagation of linear perturbations both outward and inward from the null surface \SSN\ may be of interest with regard to the causality question of the boundary condition (\rf{DN/SN}a,b) on the null surface \SSN\ in the force-free magnetosphere \citep{uch97c,uch97d,pun03,pun08}. 

\newcommand{\zetab}{\bar{\zeta}}	

\subsection{The final eigenfunctions $I(\Psi)$ and $\OmF(\Psi)$ in the force-free magnetosphere}  \label{Feigenv}
From equations\ (\rf{Iout/in}a,b) and (\rf{DN/SN}a,b), we have \citep{oka15a}
\begin{subequations} \begin{eqnarray}
\OmF(\Psi)=\omN= \frac{\OmH}{1+\zeta}, \hspace{1cm}	\lb{EigenOmFI}      \\  
I=\Iout=\Iin=- \IinU=\frac{\OmH}{2(1+\zeta)} (\Bp\vp^2)_{\rm ffH},  \hspace{1cm}	 \lb{EigenOmF}  \\  
\zeta(\Psi) \equiv (\Bp\vp^2)_{{\rm ff}\infty}/(\Bp\vp^2)_{\rm ffH}. \hspace{1cm}   \lb{Zeta}    
\end{eqnarray}  \lb{FL-eigen}   \end{subequations} 
In this eigenstate, the null surface \SSN$=$\SZAM\ will be the magneto-centrifugal plasma-shed, from which the angular momentum and the Poynting fluxes, positive and negative, flow out both ways toward $\Sffinf$ and $\SffH$. Their related AVs are given by $(\OmFm)_{\rm out}=\OmF$ in the outer domain $\calDout$, and by $(\OmFm)_{\rm in}=-(\OmH-\OmF)$ in the inner domain $\calDout$. The difference $\OmH$ of the two AVs corresponds to the voltage drop between a pair of batteries $\Dl V$ in equation\ (\rf{DV}), and this drop will lead to sustainable particle production. 

Constraints $\PG{\vcj}=\PG{\Bt}=\PG{I}=\PG{\vcv}=0$ in equation\ (\rf{EqSG}) imply that no transport of angular momentum and energy is possible within the ZAM-Gap, i.e., $\PG{\vcSJ}=\PG{\vcSE}=0$. These indicate a disconnection of current- and stream-lines between the two force-free domains and hence indicate the necessity of the current-particle sources and related EMFs in the Gap.
It will be ensured in equation\ (\rf{DN/SN}) that the copious charged ZAM-particles pair-produced in $|\OmFm|\lo\Dlom$ serve to connect and equate both $\Iout$ and $\IinU$ across the Gap $\SG$, despite $\PG{\vcv}=\PG{\vcj}=0$.
Also, the overall flow of energy-angular momentum is continuous across the ZAM-Gap as seen in equations\ (\rf{DG/SE/SJ}), regardless of $\PG{\calPE}=\PG{\calPJ}=0$ as far as the boundary condition $\DG{I}=0$ in equation\ (\rf{DN/SN}) is satisfied.  

The eigen-efficiency of extraction is given from equation\ (\rf{EigenOmFI}) by 
    \beeq 
    \epsGTE=\frac{\OmF}{\OmH}=\frac{1}{1+\zeta}
    \lb{eps} \eneq 
(see section\ \ref{eff-EE}). When the plausible field configuration allows us to put $\zeta\approx 1$ and hence $\epsGTE\approx 0.5$, we have from equations (\rf{EigenOmFI}), 
    \begin{subequations} \begin{eqnarray}
    \OmF\approx \OmFb \approx \frac12\OmH,  \lb{matching} \\  
    \hspace{1.5cm}	  c^2 |dM| \approx  \Th dS \approx \frac12 \OmH |dJ|  \lb{therfirst-eig} 
    \end{eqnarray} \lb{eig/state} \end{subequations}   
by equations\ (\rf{Laws-(ii)}a,b). 

A pair of batteries' EMFs become for $\zeta \approx 1$ by equations\ (\rf{EMF-ab}) 
    \begin{subequations} \begin{eqnarray} 
        \calEout  =-\frac{\OmH}{2\pi c} \int_{\Psi_1}^{\Psi_2} \frac{1}{1+\zeta} d\Psi 
        \approx - \frac{\OmH \Dl\Psi}{4\pi c} 	\approx -\frac{\Dl V}{2},   \hspace{0.98cm}
    \lb{eigEMF-out} 	 \\ 
    \calEin	
    =\frac{\OmH}{2\pi c}\int_{\Psi_1}^{\Psi_2}  \frac{\zeta}{1+\zeta} d\Psi \approx   \frac{\OmH \Dl\Psi}{4\pi c}	 \approx  \frac{\Dl V}{2}.   \hspace{0.7cm}  
        \lb{eigEMF-in} \hspace{0.34cm} 
\end{eqnarray}   \lb{eigEMF-ab}  \end{subequations}   
Thus, the impedance matching between particle acceleration and entropy production (MT82; TPM86)
yields good agreement with the result of the eigen-magnetosphere for $\zeta\approx 1$.

\begin{figure*}
\begin{center}
\includegraphics[width=12cm, height = 7cm, angle=-0]{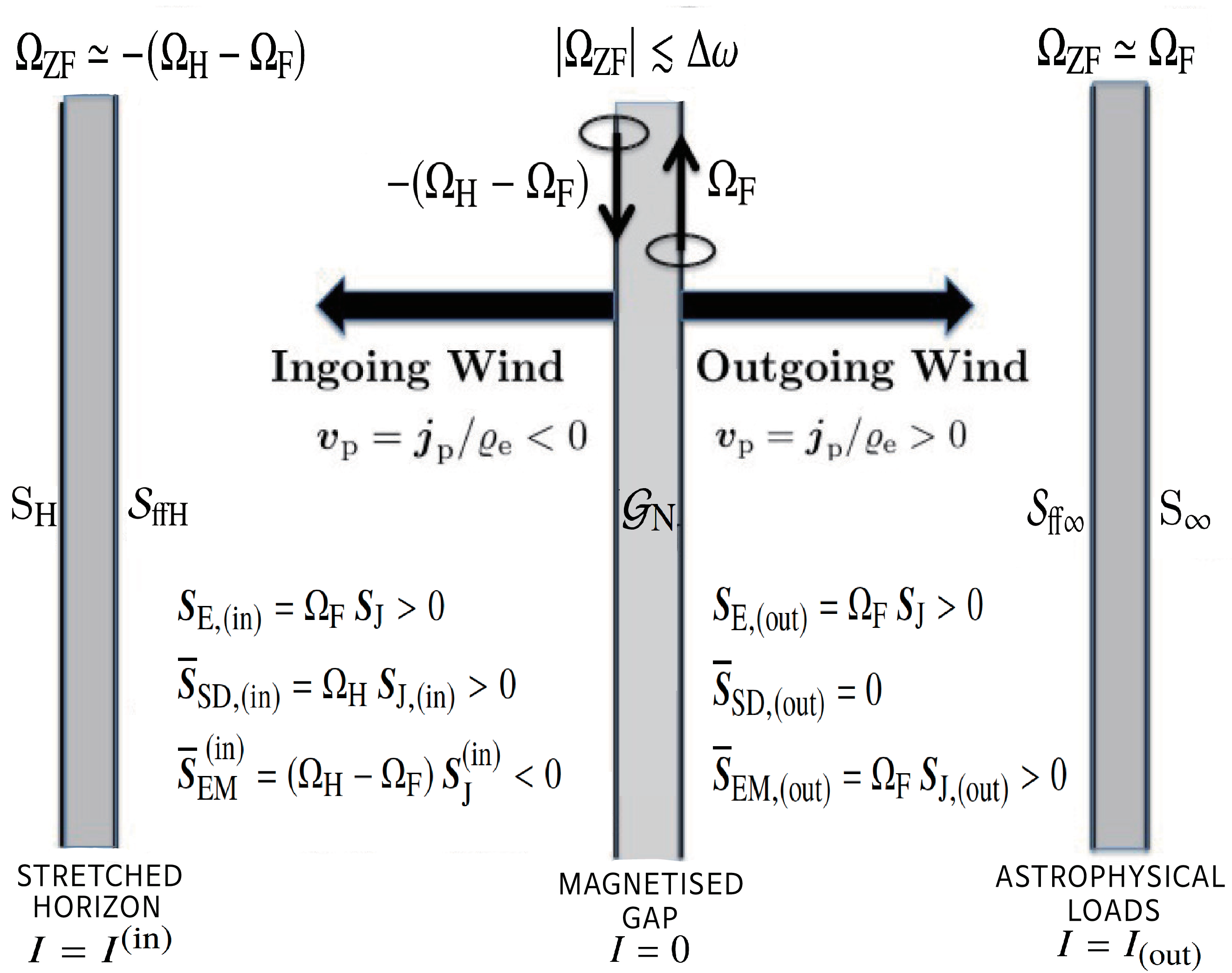}
\end{center}
\caption{A twin-pulsar model with three (two \emph{resistive} and one \emph{inductive}) membranes with $\OLOmFm$ in equation\ (\rf{OLOmFm}).  The two domains are anti-symmetric to each other with respect to the ZAM-surface \SZAM$=$ \SSN; the outer one $\calDout$ behaves like a normal pulsar-type magnetosphere rotating with the FLAV $(\OLOmFm)_{\rm (out)}=\OmF$, whereas the inner one $\calDin$ does like an anti-pulsar-type magnetosphere counter-rotating with the FLAV $(\OLOmFm)^{\rm (in)}= -(\OmH-\OmF)$. The inductive membrane $\SN$ covers the magnetized ZAM-Gap in $|\OmFm| \lo\Dlom$ at $\PG{\OLOmFm}=0$.  
The ingoing Poynting flux of negative energy in $\calDin$ from the Gap is equivalent to the outgoing Poynting flux of positive energy from the hole, just as the ingoing flux of negative angular momentum is so to the outgoing flux of positive angular momentum.  A steady pair-production mechanism due to the voltage drop $\Dl V$ will be at work to supply ZAM-particles dense enough to anchor threading field lines, thereby ensuring $\OmF=\omN$. The two batteries will provide electricity to `external resistances', where Joule heating implies particle acceleration and entropy production, respectively, in the resistive membranes $\Sffinf$ and $\SffH$ (see section\ \ref{annh}). 
}   \lbf{F-WS}  \end{figure*}

\section{A twin-pulsar model with rotational-tangential discontinuity}	   \label{TW-P-M}   
\setcounter{equation}{0}
\numberwithin{equation}{section}

Path integrals of $\vcEp$ in equation\ (\rf{EMF-ab}a,b) along the two closed circuits $\calCout$ and $\calCin$ reveal a sharp potential drop $\Dl V$ between the EMFs for the two circuits in the SC and GR domains as seen in equation\ (\rf{DV}). 
This drop comes from discontinuity RTD due to the differential rotation of the outer prograde-rotating domain with $\OmF$ and the inner retrograde-rotating one with $-(\OmH-\OmF)$ (see equation\ (\rf{DOmFm})) and will differ distinctly from any ordinary tangential and/or rotational discontinuities in classical magnetohydrodynamics (see, e.g.,\ \citet[\S 71]{lan84}). 

We attempt to briefly analyze a fundamental feature of this RTD on this surface \SSN\ in the force-free limit.  Important is the evidence that the above results, including the voltage drop $\Dl V$ across \SSN\, comes from the `continuous' function of the FDAV $\om$, and yet all these results seem also to be obtainable by assuming a `discontinuous' step function $\OLom$ for the FDAV $\om$;  
    \beeq
    \OLom= \left\{
        \begin{array} {ll}
        0 & ;\ \calDout\ (\OmFm>0), \\[1mm]   
        \omN \equiv \OmF   & ;\ \ \ \  \mbox{\rm \SSN}\ \ \ (\OmFm=0), \\[1mm] 
        \OmH  &;\ \calDin\ (\OmFm<0 ), 
    \end{array}  \right. \lb{OLom}    \eneq 
which means that $\om\approx 0$ in the SC domain and $\approx \OmH$ in the GR domain, but $\om=\omN=\OmF$ on the null surface \SSN, i.e., the ZAM-dividing surface \SZAM. 
Likewise, $\OmFm$, $\vF$ and $\epsJ$ are also replaced by the following step-functions ($\OLOmFm\equiv \OmF-\OLom$); 	
\beeq
\OLOmFm= \left\{
\begin{array} {ll}
\OmF \equiv (\OLOmFm)_{\rm (out)} & ; {\Uparrow}\ \calDout\ (\OmFm>0),	\\[1mm] 
0 \equiv (\OLOmFm)_{\rm (N)}  & ;\ \ \ \ \ \mbox{\rm \SSN}\ \hspace{3mm} (\OmFm=0),  	  \\[1mm] 
-(\OmH- \OmF) \equiv (\OLOmFm)^{\rm (in)}  &;\ {\Downarrow}\ \calDin\ (\OmFm<0), 
\end{array} \right. \lb{OLOmFm}    \eneq 
$\OLvF=\OLOmFm \vp/ \al$ and $\OLvarepsJ$ $=\OLvF(\vp \Bp^2 /c)$, 
where ${\Uparrow}$ and ${\Downarrow}$ show that
the $\calDout$ prograde-rotates with $\OmF$, while the $\calDin$ retrograde-rotates with $-(\OmH-\OmF)$, respectively (see the two arrows in figures \rff{DC-C}, \rff{GapI}, and \rff{F-WS}).  
The differences of $\OLOmFm$ and the EMFs across \SSN\ become  
\begin{subequations} \begin{eqnarray}
\DN{\OLOmFm}=(\OLOmFm)_{\rm (out)}- (\OLOmFm)^{\rm (in)}=\OmH,  \lb{DN-OmFm} \\ 
\DN{\calE}=- \frac{ \DN{\OLOmFm}}{2\pi c}\Dl \Psi=- \Dl V.     \lb{Dl-V2} 
\end{eqnarray} \lb{OmF/calE} \end{subequations}  
	
The related electric field $\OLEp$ and its discontinuity at \SSN\ become from equation\ (\rf{ZAMO-Ep})
\begin{subequations}  \begin{eqnarray}
\OLEp=-  \frac{\OLOmFm}{2\pi\al c}\vcnb\Psi, \hspace{1.0cm}     \lb{OLEp} \\  
\hspace{0.5cm}	\DN{\OLEp}=-  \frac{\DN{\OLOmFm}}{2\pi c} \LPfrac{\vcnb\Psi}{\al}_{\rm N} 
= -  \frac{\OmH}{2\pi c} \LPfrac{\vcnb\Psi}{\al}_{\rm N}.  \hspace{1.0cm}	\lb{DNOLEp}  
\end{eqnarray}  \lb{OLEpDN} \end{subequations}  
Equation (\rf{OLEp}) for $\OLEp$ naturally reproduces the same results for $\calEout$ and $\calEin$ in Faraday path integrals of $\vcEp$ along the circuits $\calCout$ and $\calCin$, respectively, as given in equations\ (\rf{EMF-ab}a,b). Also, the discontinuity of the EMF is given already in equation\ (\rf{Dl-V2}). 

The energy fluxes $\vcSEM$ and $\vcSsd$ are also replaced with the step-functions $\OLvcSEM$ and $\OLvcSsd$, respectively, i.e., 
\beeq		
\OLvcSEM=\OLOmFm \vcSJ, \quad \OLvcSsd =\barom \vcSJ  
\lb{OLvcSEMsd} \eneq  
(cf.\ $\vcSEMout$ and $\vcSEMinU$ in figure \rff{Flux-om}). 
There is naturally no discontinuity in the overall energy and angular momentum fluxes across \SSN\ with $\DN{\vcBp}=0$, i.e., similarly to equation\ (\rf{DG/SE/SJ})
\beeq
\DN{\vcSE}=\DN{\OLvcSEM+\OLvcSsd}=\OmF \DN{\vcSJ}=0. 
\lb{DGvcSE} \eneq 
We compute likewise the differences of the Poynting flux $\OLvcSEM$ and the spin-down flux $\OLvcSsd$ across \SSN\ 
from equations\ (\rf{DN-OmFm}) and (\rf{OLvcSEMsd}); 
\begin{subequations}  \begin{eqnarray}
\DN{\OLvcSEM}=  \OLvcSEMout - \OLvcSEMinU,  \lb{energyN}  \\ 
\DN{\OLvcSsd}  = - \OLvcSsdin,     \lb{energyG}  \ \  
\end{eqnarray}  \lb{vcSEMsd} \end{subequations}  
where $\OLvcSsdout=0$ and $\OLvcSEMout= \vcS_{\rm E,(out)}$, because $\om$ is regarded as negligible in the outer SC domain. Then, by equation\ (\rf{DGvcSE}), i.e.\  $\DN{\OLvcSEM+\OLvcSsd}=0$, we have 
\beeq   		
\OLvcSEMinU+\OLvcSsdin=\OLvcSEMout= \OmF\vcSJout = \OmF\vcSJin,        \lb{energyH}  
\eneq   
which is equivalent to equation\ (\rf{vcSE-c}). So, the `overall' energy flux becomes
\beeq 
\vcSEin= \OLvcSEMinU+\OLvcSsdin=-\OmF\vcSJinU= \OmF\vcSJin>0,
\lb{Totalin} \eneq  
which will be equal to $\vcSEout=\OmF\vcSJout$ across the Gap (see equations\ (\rf{energyH}) and (\rf{DG/SE/SJ}); section\ \ref{Energetic}). 

In the pulsar force-free magnetosphere, the conserved energy flux $\vcSE=\OmF\vcSJ$ alone flows outward from \SNS\ to \Sinf. In contrast, in the hole's force-free magnetosphere, the ZAMOs will see that the FD effect $\om$ forcibly split the \emph{conserved} energy flux $\vcSE$ into the two \emph{non-conserved} fluxes $\vcSEM$ and $\vcSsd$ in $\calDin$, to comply with the 1st and 2nd laws of thermodynamics. But these energy fluxes flow along the same equipotential FCSLs so that the Kerr hole will be unable to discriminate between the sum of $\vcSEM= - (\om-\OmF) \vcSJ$ and $\vcSsd=\om\vcSJ$ and that of $\OLvcSEM=-(\OmH-\OmF)\vcSJ$ and $\OLvcSsd=\OmH\vcSJ$ in the inner GR domain, while $\OLvcSEM=\vcSE$ and $\OLvcSsd= 0$ hold in the outer SC domain (see equations\  (\rf{OLom}), (\rf{OLOmFm}) and (\rf{OLvcSEMsd})). Therefore, the basic properties of the energy fluxes in the curved space with $\om$ and $\OmFm(\om;\Psi)$ will be fully taken over into the pseudo-flat space with $\OLom$ and $\OLOmFm(\OLom,\Psi)$ (see figure \rff{F-WS}; \cite{oka15a}). 

It is conjectured in the above that a kind of inevitable relaxation of the RTD due to the pair-creation by the voltage drop $\Dl V$ will lead to widening from the ZAM-surface to a ZAM-Gap $\SG$ with a finite thickness. 
This Gap may be regarded as effectively consisting of two halves of `fictitious' magnetized NSs of 'shell-like structures', e.g., the outer one forward-rotating with $(\OmFm)_{\infty}=\OmF\approx (\OmH/2)$ and the inner one backward-rotating with $(\OmFm)_{\rm H}= -(\OmH-\OmF)\approx -(\OmH/2)$, and yet the two structures are reversely packed together, and threaded by the poloidal field $\PG{\vcBp} \neq 0$ with no toroidal component and pinned down in the ZAM-particles pair-produced in the Gap, to ensure $\OmF=\om_{\rm G}$.


\section{Energetics and structure in the twin-pulsar model}  \label{FluxC}   
\setcounter{equation}{0}
\numberwithin{equation}{section} 

\newcommand{\calPEMin}{\calP_{\rm EM,(in)}}
\subsection{Energetics of the hole's extraction of energy}  \label{Energetic}   
A variant of the first law, $\OmH |dJ|=\Th dS+c^2 |dM|$, seems to indicate that the energy extracted through the spin-down energy flux will be shared at the inductive membrane $\SN$ between the two Poynting fluxes toward the two resistive membranes $\SffH$ and $\Sffinf$.  Actually, by integrating equation\ (\rf{energyH}) over all the open field lines from $\Psi_0$ to $\Psib$, we have  
\beeq	
\int_{\Psi_0}^{\Psib} \al \OLvcSsdin  \cdot d\vcA 	
= -  \int_{\Psi_0}^{\Psib} 
\alpha \OLvcSEMinU \cdot d\vcA  + \int_{\Psi_0}^{\Psib}  
\alpha \OLvcSEMout \cdot d\vcA ,  
\lb{SDenergy} \eneq   
which explains that the power $\OmH\calPJ$ extracted from the horizon is allocated between entropy production in $\SffH$ and particle acceleration in $\Sffinf$ (cf.\ MT82, Section 7.3; TPM86, Ch.\ IV D).
The two terms of the right-hand side of equation\ (\rf{SDenergy}) become, by equations\ (\rf{Iout/in}a,b), (\rf{calPEout/in}) and (\rf{ThdS/dt}), respectively;  
\begin{subequations} \begin{eqnarray} 
\TTH \frac{dS}{dt} 
=\frac{1}{2 c}   \int_{\Psi_0}^{\Psib} 	  
(\OmH-\OmF)^2(\Bp\vp^2)_{\rm ffH} d\Psi,	 \hspace{0.5cm}	\lb{H/resistanc}   \ 
 \end{eqnarray}  	
and	
    \begin{eqnarray}
    -c^2 \dr{M}{t} = \calPEout
    = \frac{1}{2c}   \int_{\Psi_0}^{\Psib}    
    \OmF^2 (\Bp\vp^2)_{{\rm ff} \infty} d\Psi \hspace{1.3cm}  \ \  \lb{L/res/a} \\  				
     =\calPEin 
    = \frac{1}{2c}  \int_{\Psi_0}^{\Psib} 	
    \OmF(\OmH-\OmF) (\Bp\vp^2)_{{\rm ffH}} d\Psi \ \hspace{0.9cm}   \lb{L/res/b} 			
    \end{eqnarray}  \lb{HL/resistance}   \end{subequations} 					
(see equation\ (\rf{Sffinf-M})), where $c^2 dM=\OmFb dJ= -\calPEin dt= -\calPEout dt$.  The left hand-side of equation\ (\rf{SDenergy}) 
reduces by equations\ (\rf{calPJout/in}) and (\rf{Iout/in})  to 
    \begin{subequations} \begin{eqnarray}
  - \OmH\dr{J}{t}  =\OmH\calPJout= \frac{\OmH}{2c}  \int_{\Psi_0}^{\Psib} 
    \OmF (\Bp\vp^2)_{{\rm ff} \infty} d\Psi \hspace{5mm}	 \lb{J/res/a} \\  
   =\OmH \calPJin = \frac{\OmH}{2c} \int_{\Psi_0}^{\Psib} 	
    (\OmH-\OmF) (\Bp\vp^2)_{{\rm ffH}} d\Psi, \ \hspace{4mm} 	 \lb{J/res/b} 	 
      \end{eqnarray}  \lb{HJ/resistance}   \end{subequations} 
(see equation\ (27) in \citep{bla22}). Summing up equations\ (\rf{H/resistanc}) and (\rf{L/res/a}) or (\rf{L/res/b}) with use of the boundary condition $\Iout=\Iin$ yields $-\OmH (dJ/dt)=\OmH\calPJ$. Also, equations\ (\rf{HL/resistance}) and (\rf{HJ/resistance}) show $\DG{\calPE}=\DG{\calPJ}=0$, despite of the RTD of the EMFs $\DG{\calE}=-\Dl V$ existent  in the Gap $\GN$ (see equation\ (\rf{DV})). 

The point is that the ZAM-particles created inside the Gap $\GN$ are spinning with $\omN=\OmF$ dragged by the hole's rotation, literally with no angular momentum. This means that the particles will easily flow out of the Gap, flung outwards or inwards from the surfaces \SgapO\ or \SgapI\ on the `plasma-shed', with positive or negative angular momenta by the respective magneto-centrifugal forces, thus keeping the ZAM-state of the Gap. 
This corresponds to the situation where the outgoing Poynting flux $\vcSEM> 0$ is related to the outer EMF $\calEout$, whereas the ingoing Poynting flux $\vcSEM<0$ is associated with the inner EMF $\calEin$.  Then, the distant observers may think as if the spin-down energy extracted through the resistive horizon membrane $\SffH$ were shared between the out- and in-going Poynting fluxes reaching the two resistive membranes $\Sffinf$ and $\SffH$, respectively, to dissipate in particle acceleration and entropy generation as seen in equation\ (\rf{SDenergy}). It seems that Kerr holes can play two roles: an acceptor of {\em negative} angular momentum and an emitter of {\em positive} angular momentum simultaneously and consistently. 


\subsection{The stream equation for the twin-pulsar model }	 \label{streamEq}  
We derive two expressions for $\jt$ ; firstly, from equations (\rf{vcj-t},c) 

\begin{subequations}	
\begin{eqnarray}	
\jt= -\frac{\OmFm\vp}{8\pi^2 \al c}\vcnb\cdot \left(\frac{\OmFm}{\al}\vcnb\Psi\right) +\frac{1}{\vp\al^2 c}\dr{I^2}{\Psi} , 
\lb{jtb} \end{eqnarray}  
and secondly, from equation (5.6b) in MT82 
 \begin{eqnarray} 
	 \jt = - \frac{\vp c}{8\pi^2\al} \left[\vcnb\cdot\left(\frac{\al}{\vp^2}\vcnb\Psi\right)+\frac{\OmFm}{\al c^2}(\vcnb\Psi\cdot\vcnb)\om \right]. 
        \lb{jt-Farad}  
 \end{eqnarray} \lb{twojt}   
 \end{subequations}    
Equating the above two expressions for $\jt$ leads to the `stream equation' in the force-free limit for the FCS-line structure in terms of the FLAV $\OmF(\Psi)$ and the ZAMO-FLAV $\OmFm$ as well as the current/angular-momentum function $I(\Psi)$;  
\beeq 
\nabla\cdot\left\{ \frac{\al}{\vp^2} \left[1-\frac{\OmFm^2\vp^2}{\al^2 c^2} \right] \nabla\Psi \right\} \nonumber 
\eneq 
\beeq 
+ \frac{\OmFm}{\al c^2}\dr{\OmF}{\Psi}(\nabla\Psi)^2  +\frac{16\pi^2}{\al\vp^2 c^2} I\dr{I}{\Psi}=0   
\lb{stream/MT} \eneq 
(see equation (6.4) in MT82). 
This reduces to the `pulsar equation' in the flat space for $\al\to 1$ and $\om\to 0$ \citep{oka74}, whereas this contains not only the two light surfaces \SOL\ and \SIL but also the null surface \SSN\ in-between, by $\vF=\pm c$ and $=0$.  The breakdown of the force-free and freezing-in conditions appears in a complicated form of severance of both current- and stream-lines, i.e., $\vcj=\vcv=0$, and the emergence of a spark-gap $\GN$, $I=\OmFm=0$, on the null surface \SSN\ (sections\ \ref{NullS} and \ref{m-mdGap}; figure \rff{GapI}). All these complications will allow the particle-current sources to be inserted into the force-free magnetosphere by particle production due to the voltage drop between a pair of batteries under the null surface \SSN. 

When the null surface \SSN\ develops into a gap $\GN$ with any finite width $|\OmFm|\lo\Dl\om$ where $I(\ell,\Psi)=0$, then the stream equation (\rf{stream/MT}) will relevantly be modified. We conjecture now that the poloidal component $\vcBp$ with no toroidal one, $\Bt=I(\ell, \Psi)=0$, will be robust enough to thread the particle-production Gap $\GN$ due to the voltage drop $\Dl V$, with the ZAM-state of the Gap maintained to keep circulation with $\OmF=\omN$ around the hole. Probably, this situation will be compatible with the solution of the stream equation  $\nabla\cdot ((\al/\vp^2) \nabla\Psi)=0$ for the particle production Gap within $|\OmFm|\lo \Dl \om$ where $\OmFm=I=0$.

\section{Discussion and conclusions}   \label{Dis-Con}
\setcounter{equation}{0}
\numberwithin{equation}{section}

\subsection{Astrophysical roles of frame-dragging in energy extraction}  \label{role-FD}  
The FD effect plays an indispensable role in reforming the pulsar force-free magnetosphere into the specifications suitable for the magnetosphere of a Kerr hole, in particular, to be adaptive to the 1st and 2nd laws of thermodynamics and to include the current-particle sources. The observance of the two laws demands a breakdown of the force-free condition.  But the issue, ``where does the breakdown take place in the force-free magnetosphere?'', seems to have remained almost untouched for more than four decades since BZ77, probably because the astrophysical roles of frame-dragging remain so far ill-understood (see, e.g., \cite{uch97a,pun96,bla02}).  This may explain why the topic of extracting energy from Kerr holes has so far continued to be a big challenge in the classical analytic approach rather than the modern numerical one \citep{tho17}. 

The ZAMOs circulating with $\om$ around the hole will be sure that the Kerr hole can extract energy if and only if frame-dragging is correctly taken into account. The coupling of FDAV $\om$ with FLAV $\OmF$ begins with the ZAMO-FLAV, $\OmFm=\OmF-\om$. Then, they will see for $0<\OmF<\OmH$ that the coupling necessarily leads to nesting the inner domain $\calDin$ counter-rotating ($\OmFm<0$) inside the outer domain $\calDout$. The magnetic sling-shot effect works inwardly through \SIL\ towards the hole in the former domain, oppositely to in the latter domain through \SOL, and hence the ZAMOs will see ``a sufficiently strong flux of \emph{negative} angular momentum leaving the null surface \SSN,'' and this does not contradict with ``a Poynting flux going towards the hole'' (BZ77). They will understand that the overall energy flux $\vcSE=\OmF\vcSJin$ always flows outward. It is on the null surface \SSN$=$\SZAM\ that the spin-axis of the hole's force-free magnetosphere changes from positive in $\calDout$ to negative in $\calDin$, and the Poynting flux emitted changes direction from outward to inward. Hence, the `complete' violation of freezing-in-ness and force-freeness `must' take place. 

The inner domain $\calDin$ with the negative-angular-momentum density ($\epsJ<0$) may be referred to as the `effective ergosphere' \citep{oka92} 
because FCSLs there represent not only negative-angular-momentum orbits but also {\em negative-energy orbits} in the ergosphere in the Penrose process.  
The ingoing Poynting flux entering into the horizon leads to the hole's entropy increase (see equation\ (\rf{Entro}a,c)) and instead allows the hole to lose positive energy. 

The counter-rotating inner domain will be designed so that the electrodynamical process of extraction of energy can surely obey the thermodynamic laws. The null surface \SSN\ is then the key surface where the eigen-FLAV $\OmF$ and the eigen-FDAV $\omN$ can simultaneously be determined uniquely, thereby dragging the force-free magnetosphere into circulation around the hole with the FDAV $\omN=\OmF$ 
non-adiabatically. It is recently pointed out that ``what is dragged by the Kerr hole are the ZAMOs and the compass of inertia'' \citep{cos21}.  The force-free magnetospheres circulating with $\omN=\OmF$ around the Kerr hole may also be included among them. 

Thanks to frame-dragging, ``a physical observer will see not only a Poynting flux of energy from \SSN\ entering the hole'' (BZ77), but also he will see another Poynting flux from \SSN\ outward, to transform into kinetic energy through the particle acceleration zone in the outer resistive membrane $\Sffinf$, and then to evolve to a high-energy gamma-ray jet beyond \citep{oka15b}. The force-free theory, including the `complete violation' of its force-freeness, will yield a self-consistent, unique theory for extracting energy from Kerr holes.  These ingenious actions of frame-dragging may seem to be due to a string puller manoeuvring behind the scenes, and we may so far never have seen and understood them as his real astrophysical effects.  The ZAMOs will not consider these relatively modest actions of frame-dragging as spooky, for they are the `physical observers' (BZ77) and the `fiducial observers' (TPM86).

\subsection{Concluding remarks} 


If observed large-scale high-energy $\gamma$-ray jets from AGNs really originate from quite near the event horizon of the central super-massive BHs, it appears to be plausible that these jets are a magnificent manifestation of the trinity of general relativity, thermodynamics, and electrodynamics (GTED);  precisely speaking,  frame-dragging, the first and second laws, and unipolar induction. The heart of the black hole's central engine may lie in the Gap $\GN$  between the two (outer and inner) light surfaces just above the horizon, and the embryo of a jet will be born in the Gap $\GN$ under the inductive membrane $\SN$. 	
The confirmation of this postulate awaits further illumination of the BH Gap physics.

\begin{ack}
I.O.\ thanks Professor Kip Thorne for his strong encouragement to continue this research (more than a decade ago). Also, he is grateful to O. Kaburaki for the joint work, which helped to deepen his understanding of thermodynamics significantly.  Y. S.\ thanks the National Astronomical Observatory of Japan for support and kindness during his visits. We are also thankful to Dr T.\ Jacobson for reminding us of their paper and useful comments. We appreciate the Research Feedback Team 
for providing helpful and encouraging comments with their advanced AI models on arXiv:2401.12684v2.\footnote{We have selected the title of our paper from the list of titles recommended in the link;  http://feedback.kellogg.northwestern.edu/UNCY5B.html}  
\end{ack}

\appendix 
\section{The place and shape of the null surface \SSN }  \label{SNshapAp} 
	The absolute space around a Kerr hole with mass $M$ and angular momentum per unit mass $a=J/Mc$ is described in Boyer-Lindquist coordinates as follows: 
\begin{subequations}
	\begin{eqnarray}
		ds^2=(\rho^2/\Delta) dr^2+\rho^2 d\theta^2 +\vp^2 d\phi^2, \\[1mm]	
		\rho^2\equiv r^2+a^2\cos^2 \theta,\ \Delta \equiv r^2-2GMr/c^2 +a^2,\\[1mm]	
		\Sigma^2\equiv (r^2+a^2)^2-a^2\Dl\sin^2\theta, \ \  \vp=(\Sigma/\rho)\sin\theta, \\[1mm]	
		\al=\rho\Dl^{1/2}/\Sigma . \ \  \om=2aGMr/c\Sigma^2  \lb{alom} \ \ 
	\end{eqnarray} \lb{Kmetric} \end{subequations}  
 where $\al$ is the lapse function, given by $\al=d\tau/dt$, that is, the ZAMO's clock's lapse/the lapse of global time $t$, and $\om=d\phi/dt$ is the angular velocity relative to absolute space (see equation (2.16) in MT82). 
 
 It will be the two parameters $\al$ and $\om$ that prepare the comfortable surrounding spacetime of a Kerr hole with the two hairs for its force-free magnetosphere with the two functions $\OmF(\Psi)$ and $I(\Psi)$.  

It is the final eigenvalue $\OmF(\Psi)$ that determines not only the efficiency $\epsGTE(\Psi)$ of energy extraction but the place and shape of the null surface \SSN, which hides a magnetized ZAM-Gap $\GN$ under it in the force-free limit. Some basic properties of the structure of force-free eigen-magnetospheres have already been clarified in some detail (see \cite{oka09,oka12a})
for the monopolar `exact' solution in the slow-rotation limit). For a tractable expression of FDAV $\om$, we deduce
\beeq 
\frac{\om}{\OmH}= 
\frac{(1+h^2)^2 x}{(x^2+h^2)^2-h^2(x-1)(x- h^2)\sin^2\theta}  
\lb{om-N} \eneq  
from equation\ (\rf{Kmetric}), where $x\equiv r/\rH$, $h=a/\rH$ and $\OmH=(c^3/2GM)h$. When we use $\omN=\OmH/(1+\zeta(\Psi))$ from equation\ (\rf{EigenOmFI}), the expression of $\xN=\xN(\theta)$ for the shape of \SSN\ with the parameters $h$ and $\zeta$ reduces to an algebraic equation; 
\begin{eqnarray}  
F_{\rm N}(x,\theta, \zeta;h)=(x^2+h^2)(x^2+h^2\cos^2\theta)   \nonumber   \hspace{1cm} \\ \hspace{1cm} 
-(1+h^2)[(1+h^2\cos^2\theta)+(1+h^2)\zeta]x=0  . \quad 
\lb{FNul} \end{eqnarray}    
It will be helpful to define a `mid-surface' \SSM\ with $\omM=0.5\OmH$ \citep{oka92} to examine topological features of \SSM$\approx$\SSN. When $\zeta(\Psi)\simeq 1$, we have 
\begin{eqnarray}  \hspace{0.5cm}
F_{\rm M}(x,\theta; h)=(x^2+h^2)(x^2+h^2\cos^2\theta)   \nonumber   \hspace{1cm} \\ \hspace{1.5cm} 
-(1+h^2)[(2+h^2(1+\cos^2\theta)]x=0 .
\lb{FM}   \end{eqnarray}	
For comparison, we consider the static-limit surface \SE\ as the surface limiting the ergosphere  from $g_{tt}=-(\Delta-a^2\sin^2\theta)/\rho^2=0$, 
\beeq
F_{\rm E}(x,\theta; h)=(x-1)(x- h^2)- h^2\sin^2\theta,
\lb{FE}  \eneq  
and its solution is expressed as
\beeq
\xE(\theta,h)=\frac{1}{2} \left((1+h^2) +\sqrt{(1-h^2)^2+4h^2\sin^2\theta} \right).
\lb{xE/sol} \eneq  
From equations\ (\rf{FM}) and (\rf{xE/sol}), for $h\ll 1$ we have 
    \begin{subequations}   \begin{eqnarray}    
    \xM = 2^{1/3} \left[1+\frac{h^2}{6} \left( 2(2- 2^{1/3})+(2^{1/3}-1)\sin^2\theta \right) \right], \hspace{6mm} \lb{xM}  \\ 
    \xE = 1+h^2\sin^2\theta,  \lb{xE}  \hspace{9mm}   
    \end{eqnarray}  \lb{xEM}  \end{subequations}  
while the two light surfaces, \SOL\ and \SIL, become for $h\ll 1$  (see equations\ (7.7a,b) in \citet{oka92})
\beeq  
\xoL=\frac{2}{h}\left(1-\frac{\sin\theta}{4} \right),   \ \ \ 
\xiL=1+\frac{h^2}{4}\sin^2\theta.   
 \lb{xoiL}  \eneq 
 
Then, we see that $\xiL<\xE<\xM<\xoL$. For $h\to 0$, it turns out that both of $\xiL$ and $\xE$ $\to 1$ and $\xoL\to \infty$, while $\xM\to 2^{1/3}=1.2599$. Therefore, when $\zeta\simeq 1$ and hence S$_{\rm M}$$\simeq$\SSN, \SSN\ will interestingly keep a position of $\xN\to 2^{1/3}$ above the horizon between $\xiL=\xE=1$ and $\xoL\to\infty$ (i.e., \SE$\leftarrow$\SIL$<$S$_{\rm M}$$\approx$\SSN$<$\SOL$\to$\Sinf), even for $h\to 0$.  

There is a certain surface \SSMc, which contacts with \SE\ from the outside at the equator, i.e., $\xM=\xE$. This occurs when $\hc=\sqrt{\sqrt{2}-1}=0.6436$, and then $\xM=1.3960$ at $\theta=0$ and $\xE=\xM=1+\hc^2=\sqrt{2}$ at $\theta=\pi/2$ (see Figs.\ 1 and 2 in \citet{oka09}).  

For the extreme-Kerr state with $h\to 1$, equation\ (\rf{FM}) reduces to 
\beeq
F_{\rm M}(x,\theta; 1)=(x^2+1)(x^2+\cos^2\theta)-2(3+\cos^2\theta)x=0,
\lb{FM1} \eneq	
which yields $\xM=1.6085$ for $\theta=0$ at the pole and $\xM=1.6344$, while by $F_{\rm E}(x,\pi/2; 1)=0$, we have $\xE=2$ at the equator (see figure 3 in \citet{oka92}). 

When $\zeta\simeq 1$, from the above analysis, one can read such interesting features at $\theta=\pi/2$ for $0\leq h\leq 1$ that 
\beeq
\begin{array} {ll}
1 \leq\xE(h) 
\leq 2, &  \mbox{for \SE}, \\ 
2^{1/3}=1.2599\leq\xN(h) 
\leq 1.6433, & \mbox{for \SSN}, 
\end{array}  \lb{rE/Neq}    \eneq 
and that $\xN \ggel \xE$ for $h\lleg \hc=(2^{1/2}-1)^{1/2}=0.6436$. This shows that, for $1\geq h\geq \hc$, the equatorial portion of the null surface \SSN\ lies within the ergosphere \SE, while, for $h<\hc$, the whole of the ergosphere \SE\ lies within the null surface. It turns out that the ergosphere changes from a spherical shape at $h=0$ to a spheroidal one at $h=1$, while when $\zeta\simeq 1$, the null surface keeps an almost spherical shape from $h=0$ to $h=1$. 
In any case, mechanical properties in the ergosphere appear to have no direct connection with electrodynamic properties of the null surface \SSN\ and the inner domain $\calDin$.


\end{document}